\documentclass[useAMS,usenatbib]{mn2e}

\usepackage{graphicx}
\usepackage{times}
\usepackage{color}
\usepackage{subfigure}

\def \aap{A\&A}

\def \aj{AJ}

\def \apss{Ap\&SS}

\def \apjl{ApJ}
\def \apjs{ApJS}
\def \apj{ApJ}

\def \mnras{MNRAS}

\def \nat{Nat}

\def \pasp{PASP}

\newcommand{\NaI}{\hbox{{\rm Na}{\sc \,i}}}

\newcommand{\MgII}{\hbox{{\rm Mg}{\sc \,ii}}}

\newcommand{\HI}{\hbox{{\rm H}{\sc \,i}}}

\newcommand{\Ha}{\hbox{{\rm H}$\alpha$}}

\newcommand{\mpy}{\hbox{M$_{\odot}$\,yr$^{-1}$}}

\newcommand{\msun}{\hbox{M$_{\odot}$}}

\newcommand{\NHI}{\hbox{$N_{\HI}$}}

\newcommand{\LL}{\hbox{$\lambda\lambda$}}

\newcommand{\kms}{\hbox{${\rm km\,s}^{-1}$}}
\newcommand{\nn}{\nonumber}
\newcommand{\hkpc}{\hbox{$h_{1.0}^{-1}$~kpc}}

\newcommand{\kpc}{\hbox{$h^{-1}$~kpc}}

\newcommand{\EW}{\hbox{$W_{\rm r}^{\lambda2796}$}}

\title[collimated winds]{Physical properties of galactic winds using background quasars}
\author[Bouch\'e et al.]{
\parbox[t]{\textwidth}{
N. Bouch\'e$^{1,2,3,\P}$, 
W. Hohensee$^4$, 
R. Vargas$^5$,
G. G. Kacprzak$^{6,\S}$,
C. L. Martin$^1$,  
J. Cooke$^6$,
C. W. Churchill$^{7}$
}
\vspace*{0.6cm}\\
$^1${Department of Physics, University of California, Santa Barbara, CA 93106, USA}\\
$^2${CNRS; Institut de Recherche en Astrophysique et Plan\'etologie [IRAP] de Toulouse,  14 Avenue E. Belin, F-31400 Toulouse, France}\\
$^3${Universit\'e Paul Sabatier de Toulouse; UPS-OMP; IRAP; F-31400 Toulouse, France }\\
$^4${Adolfo Camarillo High School, 4660 Mission Oaks Blvd. Camarillo, CA 93012}\\
$^5${Oak Lawn Community High School, 9400 Southwest Hwy.,
Oak Lawn, IL 60453}\\ 
$^6${Center for Astrophysics and Computing, Swinburne University of Technology, Mail H30, PO Box 218, Hawthorn, Victoria 3122, Australia}\\
$^7${Department of Astronomy, New Mexico State University, Las Cruces, NM 88003 USA}\\
$^{\P}${Marie Curie Fellow}\\
$^{\S}${Australian Research Council Super Science Fellow}\\
}

\begin{document}

\date{Accepted --- Received}

\pagerange{\pageref{firstpage}--\pageref{lastpage}}  

\maketitle

\begin{abstract}
Background quasars are potentially sensitive probes of galactic outflows   provided that one can determine the origin of the absorbing material since both
 gaseous disks and strong bipolar outflows can contribute to the absorption cross-section. 
Using a dozen quasars passing near spectroscopically identified galaxies at $z\sim0.1$, we find that the  azimuthal orientation of the quasar sight-lines
with strong \MgII\  absorption (with $\EW>0.3$\AA) is bi-modal:
about half the \MgII\ sight-lines are aligned with the major axis  and the other half are within $\alpha=30^{\circ}$ of the minor axis, suggesting that bipolar outflows can contribute to the \MgII\ cross-section. 
This bi-modality  is also  present in the instantaneous star-formation rates (SFRs) of the hosts.  
For the sight-lines aligned along the minor axis, 
a simple bi-conical wind model is indeed able to reproduce 
the observed \MgII\ kinematics and the \MgII\ dependence with impact parameter $b$, (\EW$\propto b^{-1}$). 
 Using our wind model, we
can  directly extract key wind properties such as the de-projected outflow speed $V_{\rm out}$ of the cool material traced by \MgII\ and  the outflow rates $\dot M_{\rm out}$. 
The outflow speeds $V_{\rm out}$  are found to be 150-300~\kms, i.e. of the order of the circular velocity, and smaller than the escape velocity by a factor of $\sim2$.   
The outflow rates $\dot M_{\rm out}$ are typically two to three times the instantaneous SFRs.
Our results demonstrate how background quasars can be used to measure wind properties with high precision.
\end{abstract}


\section{Introduction}

In spite of our understanding of the growth of dark-matter structures
from the initial density fluctuations 
\citep[e.g.][]{WhiteS_78a,MoH_02a},  the  halo mass function over-predicts the observed number density of galaxies both at the low- and high-mass ends of the mass function \citep[e.g.][and references therein]{CrotonD_06a,vandenBoschF_07a,ConroyC_09a,BehrooziP_10a,MosterB_10a,GuoQ_10a,FirmaniC_10a}.
This major discrepancy requires a (or several) mechanism to somehow suppress galaxy formation.
 
Super-novae (SN) driven winds are often invoked because they
could  suppress star formation in low-mass galaxies ($L<L^*$), 
\citep[e.g.][]{DekelA_86a,OppenheimerB_06a,OppenheimerB_10a}
and   transport large amounts of energy and gas out of young galaxies  and enrich the inter-galactic medium (IGM).
This scenario is supported by the fact that
 galactic winds are ubiquitous in all types of star-forming galaxies: 
in local starburst galaxies \citep[e.g.][]{LehnertM_96a,HeckmanT_00a,StricklandD_00a,StricklandD_04a,MartinC_98a,MartinC_02a,SchwartzC_06a},
in extreme starbursts, such as the Ultra Luminous Infra-Red galaxies (ULIRGs)
\citep[e.g.][]{MartinC_05a,RupkeD_05a,MartinC_06a,MartinC_09a}, and
in normal star-forming galaxies both at   intermediate  \citep{SatoT_09a,WeinerB_09a,RubinK_10a,RubinK_10b}
and high-redshifts \citep{PettiniM_02c,ShapleyA_03a,GenzelR_11a}.

Numerical simulations  must often invoke strong galactic outflows in order to reproduce the luminosity function and the enrichment of the IGM \citep[e.g.][]{OppenheimerB_06a,OppenheimerB_10a,SchayeJ_10a,WiersmaR_11a}. 
These simulations  must, however, postulate 
scaling relations for the wind speeds and 
outflow rates, etc.,  in order to reproduce observational constraints. 
For instance, Dav\'e and collaborators assume that the outflow rate, $\dot M_{\rm out}$, is proportional to the SFR ($\dot M_{\rm out}=\eta\;$SFR), where the loading factor $\eta$ is a function of halo mass.
For momentum- (or energy-) driven winds, $\eta$ is proportional to $ V_c^{-1}$ (or $V_c^{-2}$), respectively.

Unfortunately, most wind properties (e.g. the opening angle, the outflow rates, the loading factors)
are poorly constrained. The best estimates of  $\dot M_{\rm out}$ made by several groups over the past decades \citep[e.g.][]{HeckmanT_90a,HeckmanT_00a,PettiniM_02b,MartinC_02a,MartinC_05a}  using
galaxy absorption line
 spectroscopy  are usually uncertain by orders of magnitude. 
One reason for these large uncertainties is that   one must estimate
the total gas column in the wind from the ion column density which requires assumptions for the gas metallicity and the ionization factor. 
Another reason is that traditional spectroscopy 
 \citep[e.g.][]{LehnertM_96a,HeckmanT_00a,MartinC_98a,RupkeD_05a,MartinC_06a,SchwartzC_06a,WeinerB_09a,RubinK_10a,RubinK_10b} probes the wind looking `down-the-barrel',
i.e. it provides no information on the physical location of the material, as the blue-shifted material could be located at 0.1, 1 or 10 kpc from the host. 
In addition, the wind geometry is unknown and as a result the wind solid angle is often assumed to be  $\Omega_w=4\pi$. Lastly, the absorption trough could be filled with difficult-to-remove emission \citep[e.g.][]{ProchaskaJ_11a}.  Each of these factors make
estimates of outflow rates very uncertain.

The radial extent of the wind can be addressed directly with
background galaxies, as demonstrated by \citet{SteidelC_10a} at $z\sim2$,    
  and by \citet{BordoloiR_11a}  at $z\sim1$. 
However, apart from exceptional cases \citep[e.g.][]{RubinK_10a},
one must stack the spectra of dozens or hundreds of background galaxies in order to gain sufficient signal-to-noise.   This stacking inevitably leads to averages in the geometries involved \citep{SteidelC_10a}.
But when sufficiently large samples are available,
the azimuthal dependence can be revealed, as demonstrated by \citet{BordoloiR_11a} who showed that the rest-frame \MgII\ equivalent width of background galaxies is  strongest along the minor axis.

One can also use background quasars to probe the radial extent of the wind.
Background quasars have several advantages to the other techniques. For instance,
they allow us to probe gaseous material of {\it any} distant star-forming galaxy  irrespective of its luminosity \citep[as in][]{StockeJ_04a,BowenD_05a,TrippT_05b,ZychB_07a}.
Compared to background galaxies, background quasars are better  probes  because  they
require no stacking, i.e. the geometry of the absorbing flow is preserved and no loss of information occurs in azimuthal averages.
Compared to galaxy absorption line spectroscopy, they allow us to probe the material
  at a known distance from the original source.

The low ionization \MgII\ doublet (\LL2796,2803) seen in background quasars (QSOs)  is   ideal for probing galactic winds
as it can be observed from $z\sim0.1$ to $z\sim2.2$ in the optical and has been associated mostly with star-forming galaxies since
the work of \citet{BergeronJ_88a}, \citet{BergeronJ_91a}, \citet{SteidelC_92a}, \citet{SteidelC_94a}.
 Unfortunately, the physical origin of strong absorbers is still debated.
Indeed, \MgII\ absorbers  
could probe the cool ($T\sim10^4$ K) material entrained in galactic winds \citep[e.g.][]{NulsenP_98a,SchayeJ_01a,MartinC_06a,CheloucheD_10a},  
  the outskirts of  gaseous disks \citep[e.g.][]{ProchaskaJ_97b,SteidelC_02a,KacprzakG_10a},
 the halos of galaxies \citep[e.g.][]{BahcallJ_69a,MoH_96a,LanzettaK_92a,SternbergA_02a,MallerA_04a},
  infalling material \citep[e.g.][]{TinkerJ_08a,KacprzakG_10a,StewartK_11a}
or a combination of these mechanisms \citep[e.g.][]{CharltonJ_98a}.
While mounting evidence points to galactic winds for strong \MgII\ systems \citep{BondN_01a,BoucheN_06c,MenardB_10a,NestorD_11a},
a direct link between low-ionization metal lines and galactic winds 
has yet to be established.

The debate on the origin of strong \MgII\ absorbers arises because it is difficult
 to build large samples of individual quasar-galaxy
pairs.  Indeed, at low redshifts ($<0.1$),  the frequency of such pairs is low
\citep[e.g.][]{BowenD_95a,StockeJ_04a,BowenD_05a}, 
and at high-redshifts it is time consuming to identify the galaxies 
associated with QSO absorption lines.  Even though  significant samples of quasar-galaxy pairs
are available \citep[e.g.][]{ChurchillC_96a,ChenH-W_01a,NoterdaemeP_10a,ChenHW_10b,RaoS_11a,LovegroveE_11a},  
there are only 19 $z\sim0.5$--1 galaxy-quasar pairs where the host galaxy kinematics have been
compared to the absorption kinematics \citep{SteidelC_02a,EllisonS_03a,ChenH-W_05a,KacprzakG_10a},
excluding the 14 pairs of \citet{BoucheN_07a} whose analysis is in progress.

In this paper, we use the unique sample of about a dozen
 $z\sim0$ galaxy-\MgII\ pairs from
\citet{BartonE_09a} and \citet{KacprzakG_11a} in order to investigate the
 relative orientations of the quasar lines-of-sight with respect
 to the host galaxy orientation.
 Section~\ref{section:sample} summarizes the properties of the sample. 
We show in Section~\ref{section:results} that the sample is made of two classes of \MgII\ absorbers.
In Section~\ref{section:winds}, we discuss the physical properties of galactic outflows
for the sub-sample of pairs related to outflows. Finally, we discuss the implications of our results in Section~\ref{section:conclusions}.

Throughout, we use a  `737' cosmology, with $h=0.7$, $\Omega_M=0.3$, and $\Omega_\Lambda=0.7$.
 
\section{Sample }
\label{section:sample}

A recent increase  in blue sensitivity of the Keck 
 Low Resolution Imaging Spectrograph (LRIS) opened a new redshift
window and allows the detection of the \MgII\ doublet down to  $z \simeq 0.1$
as demonstrated by \citet{BartonE_09a}.  In combination with the spectroscopic completeness of
the Sloan Digital Sky Survey \citep[SDSS][]{YorkD_00a,SchneiderD_10a} at that same redshift, one has the possibility to
study large unbiased  samples of nearby \MgII--galaxy pairs.
 \citet{BartonE_09a} constructed such a sample designed 
to probe for the presence (or absence) of \MgII\ absorption in a well-understood,
 volume-limited spectroscopic survey of galaxies at $z \sim 0.11$ with luminosity $M_r\la-20.5$ ($\sim L^*$). 
Out of the 20 sight-lines passing within 75\hkpc\ of   $z\sim0.1$ luminous galaxies, six exhibit strong ($\EW>0.3$~\AA) \MgII\ absorption at the same redshift as the galaxy. 
\citet{KacprzakG_11a} extended this sample to 13 such galaxy-\MgII\ pairs   using the same observational strategy.

This is the largest sample of quasar-galaxy pairs at  $z\sim0$, which has the advantage that follow-up observations are either available in the SDSS database
 or easy to obtain using 4-m or 8-m class telescopes.
This strategy led \citet{KacprzakG_11a} to measure the rotation
curves of the hosts using the Apache Point Observatory (APO) and present  a detailed comparison between the host galaxy kinematics and the absorption kinematics.
Hence, the galaxies in this sample have reliable systemic redshifts, a key aspect for our study.

The galaxy SFRs are computed using the $\Ha$ luminosities 
 measured from SDSS spectra  using the formalism of \citet{KewleyL_02a} 
assuming a \citet{SalpeterE_55a} initial mass function (IMF) and no
intrinsic reddening~\footnote{This is equivalent to assuming a \citet{ChabrierG_03a} IMF
and $A_V=0.8$~mag of extinction, typical for star-forming galaxies.}.
Due to the small angular size of the SDSS fiber apertures,   the SFRs were scaled by the ratio of
the $r$-band galaxy total counts to those contained within the SDSS
fiber.

The sample of 13 galaxy-quasar pairs of \citet{KacprzakG_11a}
is made of 11 quasars and 13 galaxies, two of which are associated with the  same quasar line-of-sight.  In this study, we kept only the galaxy with the smallest impact parameter, and were left with 11 unique galaxy-quasar pairs.  
Table~\ref{table:summary} lists the
observed properties of the sample, taken from \citet{KacprzakG_11a}.

\section{Results}
\label{section:results}

\subsection{Two classes of absorbers}
\label{section:twoclasses}

\begin{figure}
\centering
\includegraphics[width=7cm]{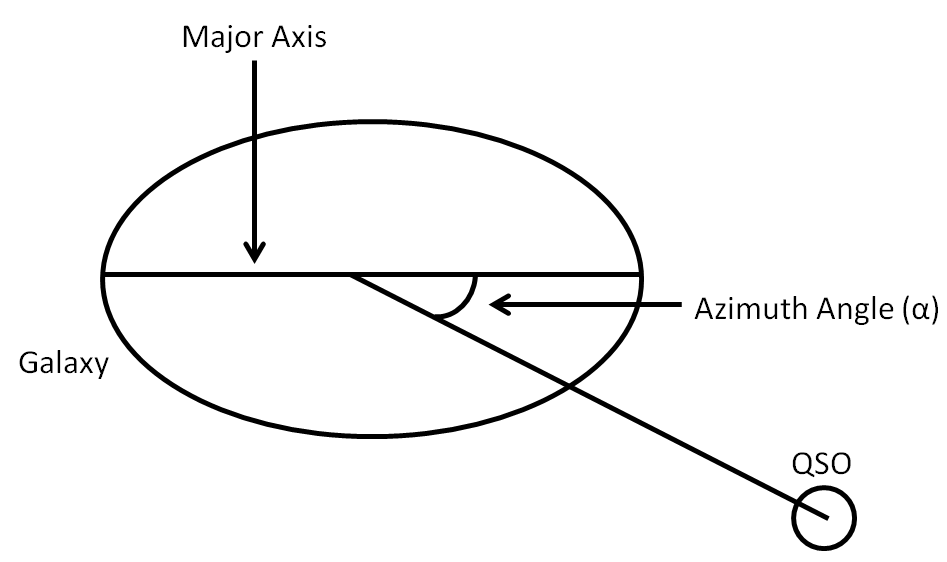}
\caption{Schematic diagram of an inclined disk, showing the relative azimuth angle $\alpha$ measured  with respect to the galaxy major-axis. }\label{fig:alpha}
\end{figure}

\begin{figure}
\centering
\includegraphics[width=9cm]{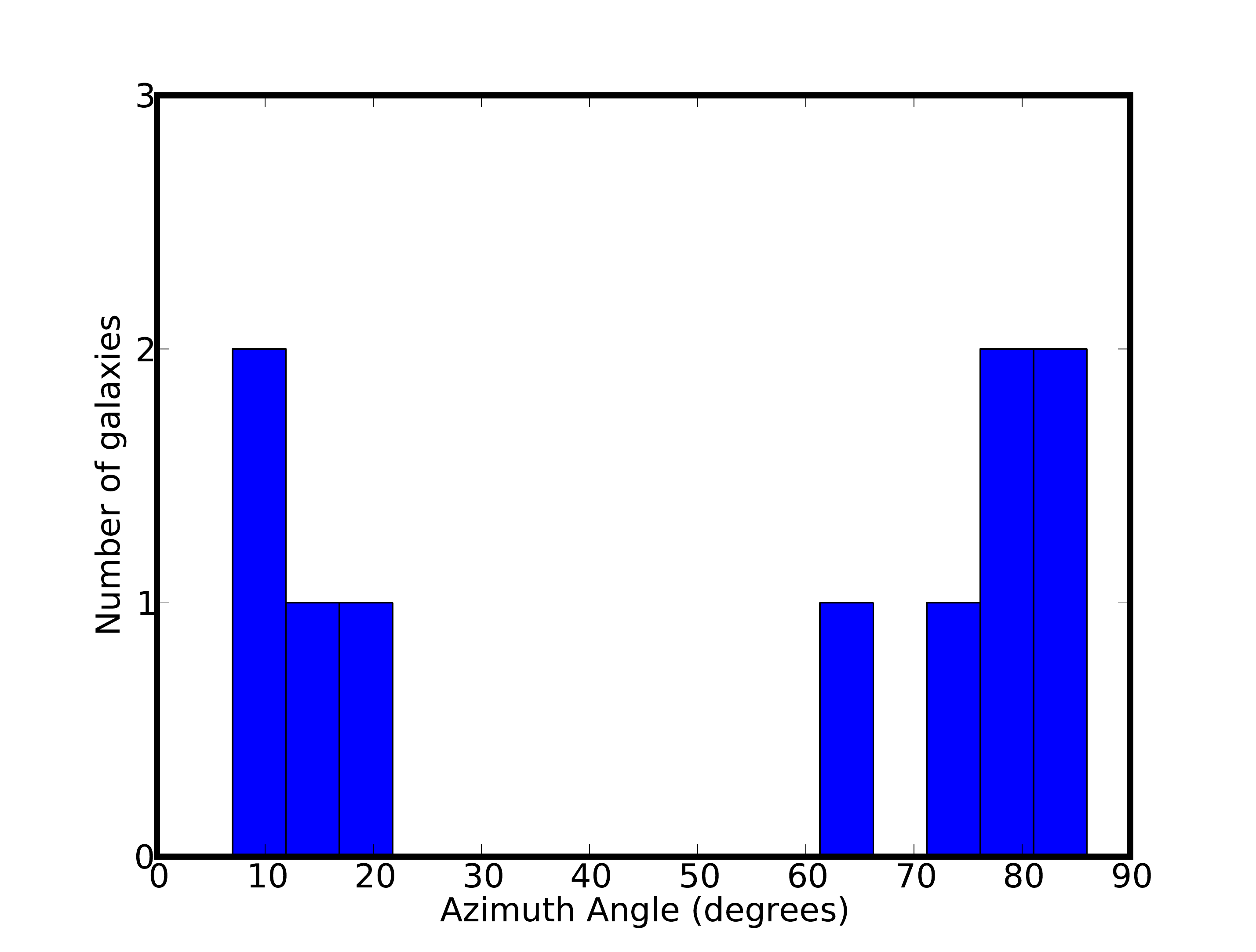}
\caption{The bimodal distribution of azimuth angle $|\alpha|$  for our sample made of 10 galaxy-quasar pairs where $\alpha$ could be determined. 
About half the sample has $|\alpha|$ less than 20~deg., and the other half
has $|\alpha|\ga60$~deg. The distribution suggests that extended gaseous disks and galactic winds contribute significantly to the \MgII\ cross-section in roughly equal proportions. 
 The lack of galaxy-quasar pairs with $|\alpha|$ between 20 and 60~deg.
is significant at the $\ga3.0\sigma$ level (see text). 
}
\label{fig:bimodal}
\end{figure}  

As mentioned in the introduction,
the physical mechanism (e.g. galactic winds, infalling gas, extended disks) that can produce \MgII\ absorption has been debated for decades.  As discussed in \citet{KacprzakG_11b},
if there are preferred  kinematics and spatial
distributions of \MgII\ absorbing gas relative to the host galaxies,
then the absorption strengths  would follow a predictable behavior as a function of galaxy orientation and/or relative orientation
to the QSO sight-line, for instance. 

Here, we investigate the distribution of the azimuth angle $\alpha$ of the quasar line-of-sight
with respect to the galaxy major axis, as illustrated in Figure~\ref{fig:alpha}.   
By examining the azimuth angle distribution, one could conclude that   galactic winds
dominate the \MgII\ cross-section if the quasars are preferentially aligned
with the minor axis ($\alpha\sim90$),  since this is the only mechanism that can produce cool material 
systematically along the minor axis. Alternatively, one could conclude
that the extended parts of gaseous disks, or infalling material \citep[according to][]{StewartK_11a} would dominate the \MgII\ cross-section if the quasars are preferentially aligned with the galaxy orbital plane \citep[e.g.][]{CharltonJ_96a}.
Furthermore,  if there is no preferred $\alpha$, then this would point towards the halo model where the gas clouds traced by \MgII\ are uniformly distributed \citep[e.g.][]{LanzettaK_92a,SteidelC_94a,MoH_96a,TinkerJ_08a}.

Looking at the relative distribution of  QSO lines-of-sight
has previously not been possible  because of  the low number of unbiased pairs 
whose kinematic axis is known.  
Indeed, to our knowledge, at $z\sim0$ there exists only one such sample, the sample of 11 galaxy-quasar pairs of \citet{KacprzakG_11a}, originally from the volume-limited survey of \citet{BartonE_09a}~\footnote{At intermediate  redshifts, there is the sample of 19 $z\sim0.5$--1 galaxy-quasar pairs from \citep{SteidelC_02a,EllisonS_03a,ChenH-W_05a,KacprzakG_10a} and the 14  pairs discovered by \citet{BoucheN_07a}.}.  

We remeasured the azimuth angles because we noticed some inconsistencies  in the  $\alpha$'s   reported
by \citet{KacprzakG_11a}. To do so, we measured the inclinations ($i$), position angles (PA) of the major axis and azimuth angles  ($\alpha$) of the quasar using two methods:
 a  visual inspection of the images and a parametric \citet{SersicJ_63a} fit to the SDSS postage stamp
images. The inclination (via its axis ratio $b/a$),  the S\'ersic index, $n$, and the galaxy PA  were fitted with  custom routines, where the S\'ersic profile is convolved with the image Point Spread Function (PSF).  We note that our inclination measurements   obtained
from visual inspection (`Manual'), from our S\'ersic fits (`Fit') and from a full bulge-disk decomposition by  \citet{KacprzakG_11a} agree well with each other.
Table~\ref{table:measurements} lists the inclination and azimuth angle measurements.

  We found that the inclinations can be reliably determined, as our fitted values are 
within 10\%\ to those derived by \citet{KacprzakG_11a}.  On the other hand, we found that our  azimuth angles  differ significantly, and we attribute the difference to a mistake in the image orientation.  We note that for one galaxy (J161940G1), its PA is poorly determined
 as it is observed almost perfectly face-on ($b/a\sim1.0$), i.e. its major-axis PA  and its azimuth angle   is undefined.  Hence, we are left with a sample of 10 galaxy-quasar pairs with reliable azimuth angles.

Figure~\ref{fig:bimodal} shows the distribution of  $|\alpha|$.  The distribution is strongly bi-modal,
with a subset made of four galaxies with small $|\alpha|$'s and another made of  six  with $|\alpha|\sim90$.  In other words, the distribution shows that all the quasar-galaxy pairs are  either nearly aligned with the major or with the minor axis.
The lack of quasar-galaxy pairs with $\alpha$ between 20 and 60~deg  is not consistent with small number statistics.  Indeed,
the probability of having no quasar-galaxy with $|\alpha|$ between $\sim$20 and $\sim$60~deg is 0.2\%\ using 10$^6$ simulated samples (with $N_{\rm pairs}=10$) drawn from a uniform distribution $U(0,90)$. 
Thus,  the central gap in the distribution shown in Fig.~\ref{fig:bimodal}
is significant at the $\ga$3.0--$\sigma$ level.

In short, we found a strong azimuthal dependence in the presence of \MgII, in good agreement with the results of \citet{BordoloiR_11a} obtained at  $z\sim1$ in the Cosmological evolution Survey (COSMOS).   
In stacked spectra of background galaxies, \citet{BordoloiR_11a} also found a strong azimuthal dependence of the total (2796\AA,2803\AA) rest-frame equivalent width $W_r$ for pairs within 40~kpc of inclined disks.
Similarly, in stacked spectra of thousands of
local star-forming galaxies from SDSS/DR7,
\citet{ChenY_10a} showed that the blue-shifted \NaI~D absorption is
stronger within 60$^{\circ}$ of the minor axis.

 Contrary to stacked spectra, the background quasars (Fig.~\ref{fig:bimodal}) reveal that 
there are two distinct populations of \MgII\ absorbers, in roughly equal proportions.   
One population of absorbers could be associated
with extended gaseous disks (or other processes aligned with the disk) and the other could be associated
with galactic winds.  At this stage, the wind scenario is only based on a plausibility argument, namely
the expectation that all the other physical processes (accretion, extended gaseous disks, halo gas) are expected to produce absorption that are not aligned with the minor axis. 
In the next section, we will investigate the wind scenario further and attempt to match
the observed \MgII\ kinematics using a simple wind model. 
This exercise will show that the classification based on $|\alpha|$ is strongly supported by our kinematic model.
Table~\ref{table:measurements} lists the classifications.

\begin{figure*}
\subfigure[]{
\includegraphics[width=14cm]{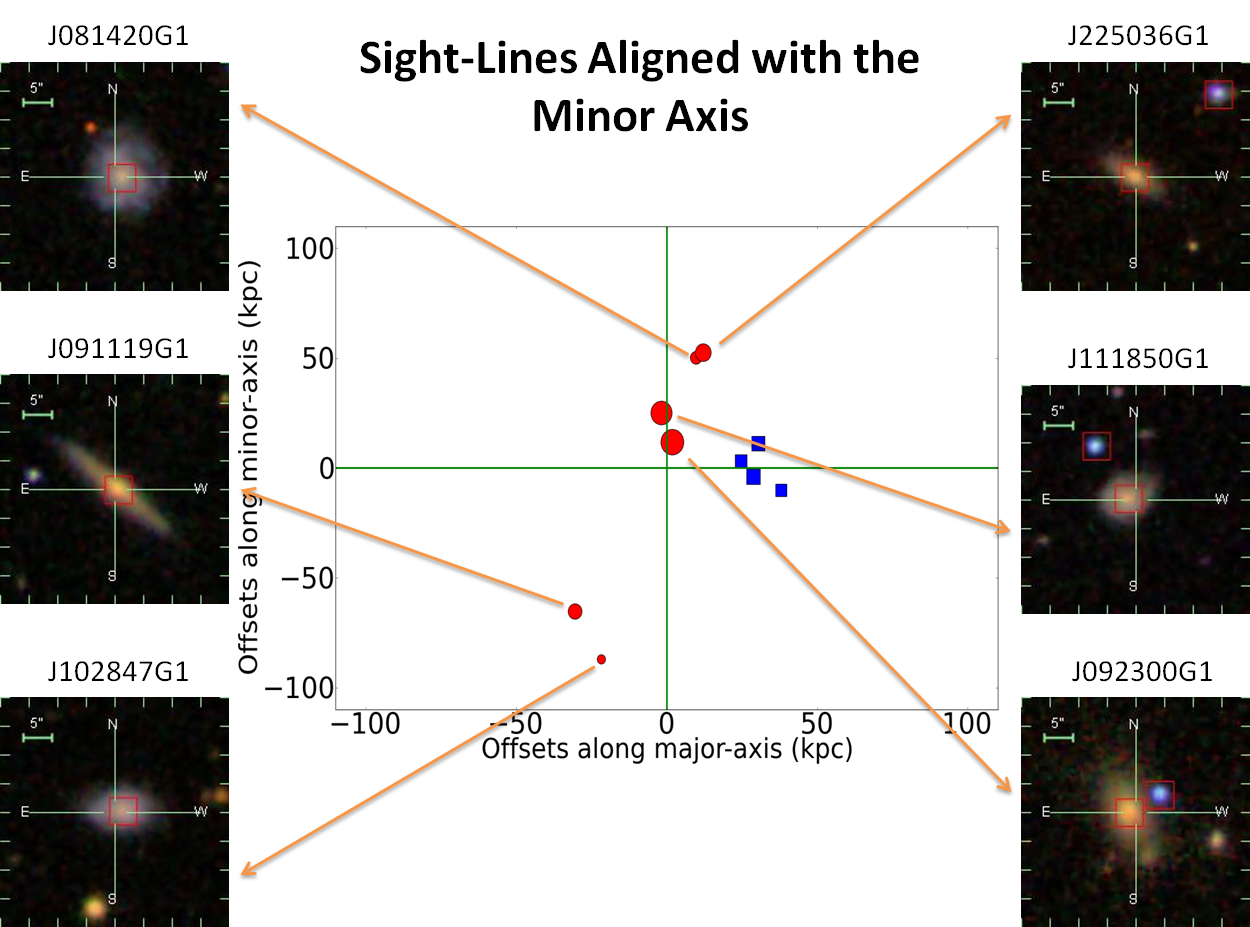}
}
\subfigure[]{
\includegraphics[width=14cm]{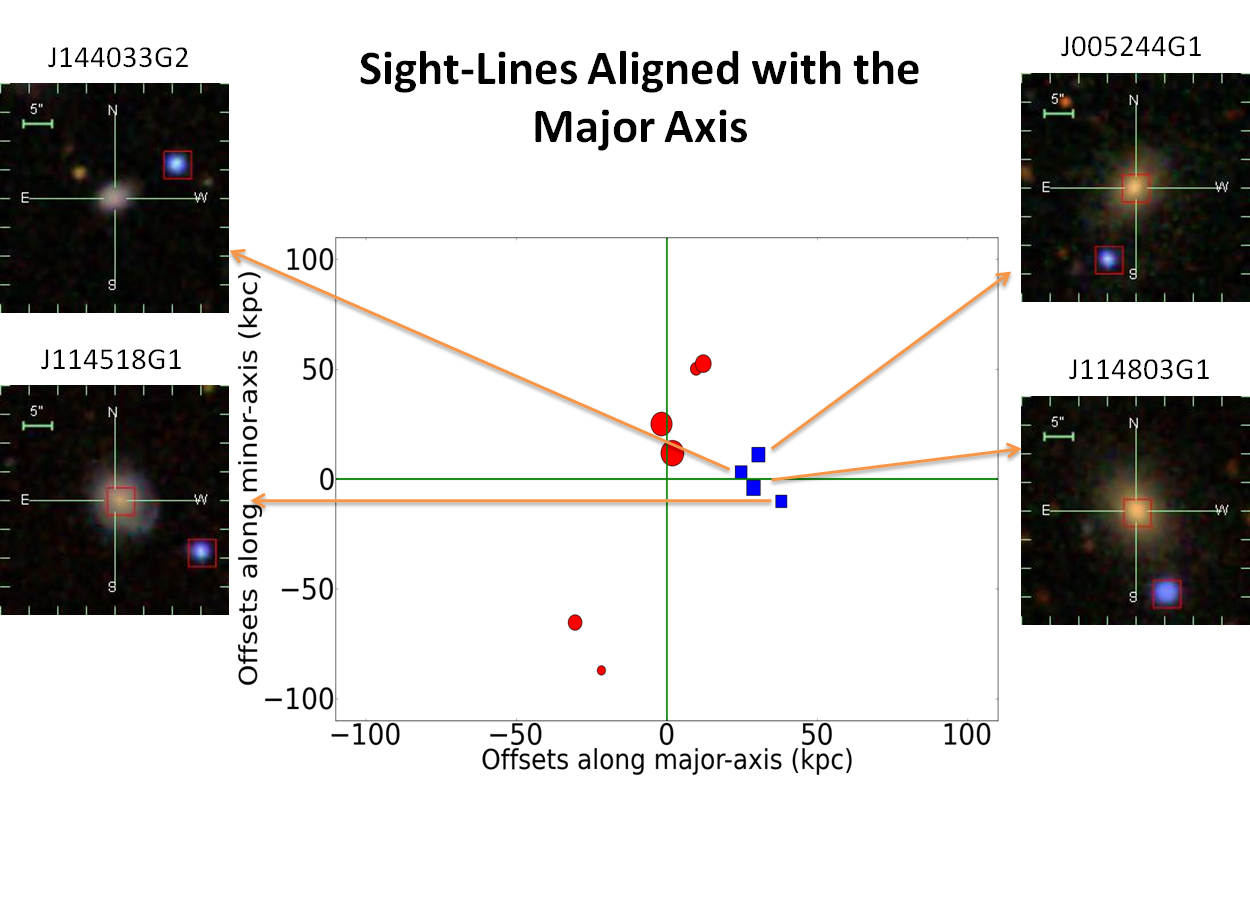}
}
\caption{Position of the quasar line-of-sight distribution relative to the galaxy major axis (x-axis) and minor-axis (y-axis) for the galaxy-quasar
pairs with high (a, circle symbols) and low (b, square symbols) azimuth angles $\alpha$. 
The size of the symbols is proportional to the rest-frame \EW.
The postage stamp images show the color SDSS image of the associated galaxy. 
Blue point source objects with squares indicate the QSO location  when visible within the field-of-view. }
\label{fig:dxdy}
\end{figure*}

\begin{figure}
\centering
\includegraphics[width=9cm]{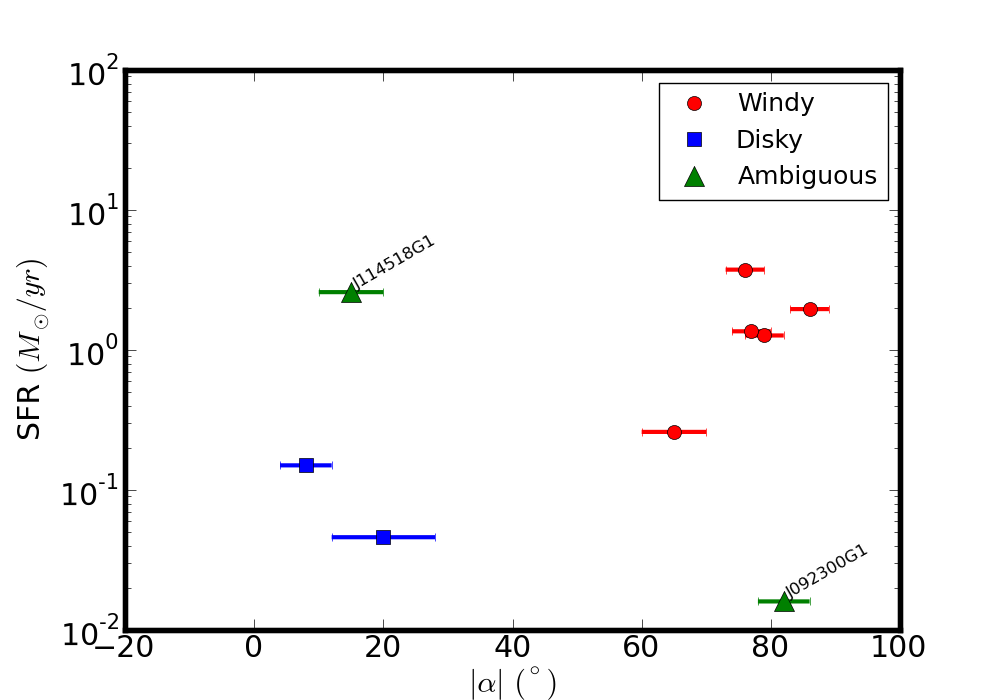}
\caption{SFR as a function of $|\alpha|$ shows that
galaxies associated with bi-conical galactic winds (high $\alpha$) have higher SFRs than those with low $\alpha$'s. SFR are computed from the nebular emission lines in the SDSS spectra and scaled to take into account the SDSS fiber size.   The kinematic modeling in section~\ref{section:indivcases} show that
two galaxies (J092300G1 and J114518G1) are misclassified based solely on the azimuth angle $\alpha$ criterion. } 
\label{fig:SFR}
\end{figure}

Having identified two classes of absorbers, we searched for differences in the galaxy properties 
between the two sub-classes.
In Figure~\ref{fig:dxdy}, we show the galaxy postage stamp images  and the quasar location relative to the galaxy major and minor axes.
 The red and blue squares show the relative QSO positions for the sub-sample with $|\alpha|>60$ and $|\alpha|<20$, respectively.  We find no significant differences in the galaxy colors  between the two sub-samples.
The most significant difference between the two is found in their SFR distribution.   Figure~\ref{fig:SFR} shows that galaxies  with high $|\alpha|$ (classified as `windy')
have higher SFRs, while  pairs with low $|\alpha|$'s (`disky') have lower SFRs. 
Note the kinematic modeling presented in section~\ref{section:indivcases} will show that
two galaxies are misclassified based solely on the azimuth angle  criterion.

\subsection{The \MgII\ kinematics of the Wind sub-sample: Wind modeling}
\label{section:model}

In light of the results presented in section~\ref{section:twoclasses},
we postulate that the \MgII\ absorption is  produced by material entrained in galactic winds for the sub-sample made of galaxy-quasar pairs 
aligned along the galaxy minor axis (with $|\alpha|\sim90$).
To test  this hypothesis, we construct 
 a simple bi-conical wind model aimed at reproducing the
 observed \MgII\ kinematics. 
Our simple galactic wind model is made of 10$^5$ `clouds'  distributed in a cone within an opening-angle $\theta_{\rm max}$ (and corresponding solid angle $\Omega_{\rm w}=\pi\theta_{\rm max}^2)$~\footnote{After some experimentation,  we found that a hollow cone with $\theta_{\rm min}\sim10^{\circ}$ performed somewhat better in reproducing the shape of the \MgII\ absorption profile. 
Note the kinematic results (absorption centroid and width) are completely insensitive to $\theta_{\rm min}$.}.
The discrete  clouds populate the cone  from a minimum radius $R_{\rm min}=5$ kpc to a maximum radius $R_{\rm max}\simeq100$~kpc, covering the range of impact parameters.
We assume that the clouds are entrained in the bi-conical wind and
  are moving at a {\it constant} velocity ($V_{\rm out}$), which
  is the only model parameter fitted against the data.

All the parameters related to the geometry of the wind  can be
determined from the data. The wind opening angle is  $\theta_{\rm max}\sim30$~deg 
according to the distribution of azimuth angles (Fig.~\ref{fig:bimodal})
since no QSO-galaxy pairs are found beyond $\pm30$~deg from the minor axis. 
 The corresponding  solid angle is thus  of order unity with
 $\Omega_{\rm w}\simeq0.86$.
Similarly, the relative geometric orientation of the wind
with respect to the quasar line-of-sight is also given by the data.
 For all galaxy-quasar pairs, the galaxy inclination ($i$) and the relative orientation ($\alpha$) of the sight-line are set by the data.
The only degree of freedom left is to choose 
 whether the cone intercepted by the quasar sight-line
 is pointing either towards or away the observer. 
We adopt the convention that  $ x, y$ are the coordinates in the plane of the sky, with $ x$ along the galaxy major axis and
consequently, the $z$-axis  along the quasar line-of-sight.

\subsubsection{Notes on individual cases}
\label{section:indivcases}

\begin{figure*}
\subfigure[]{
\includegraphics[width=11cm]{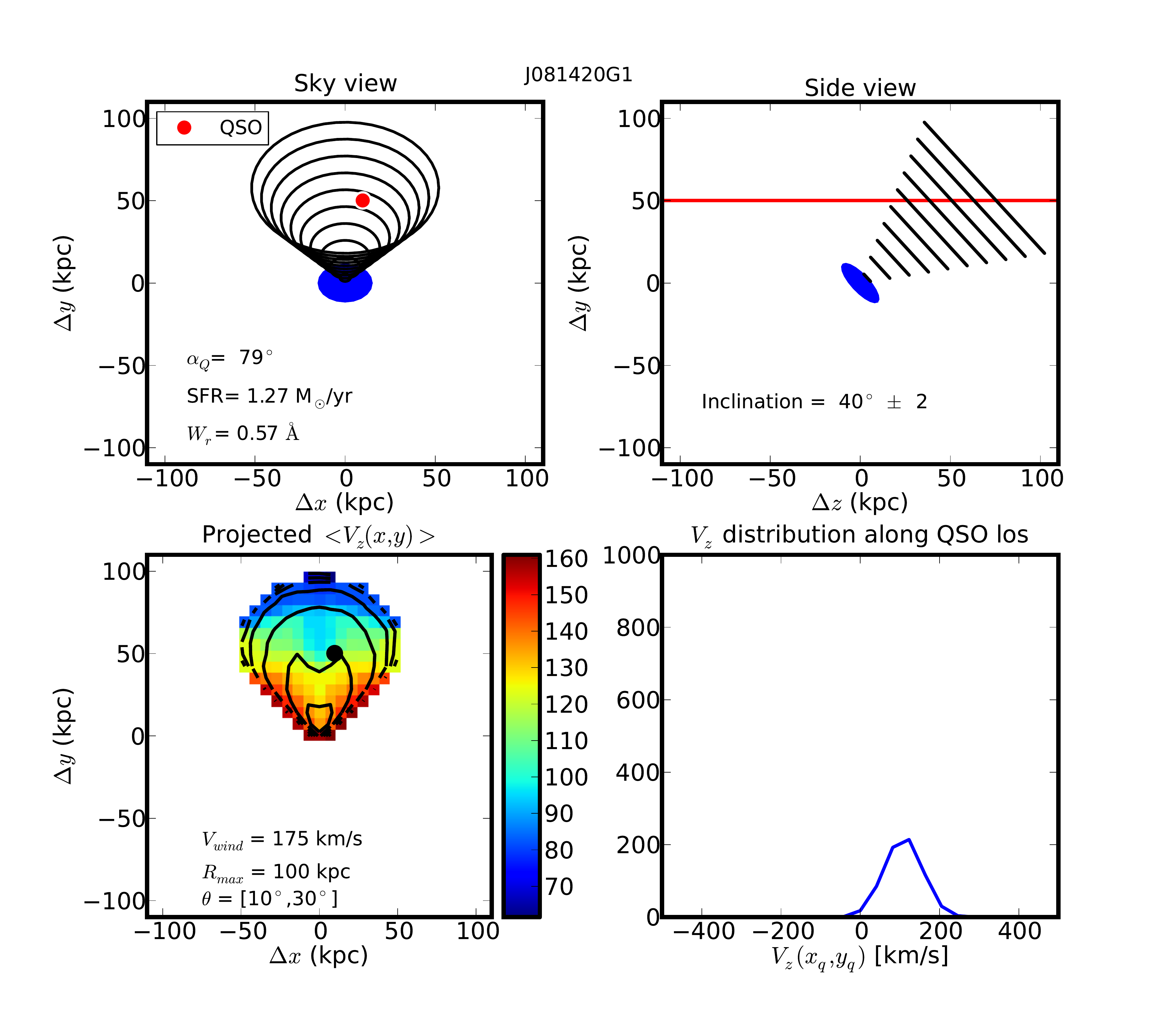} 
}
\subfigure[]{\includegraphics[width=5.5cm]{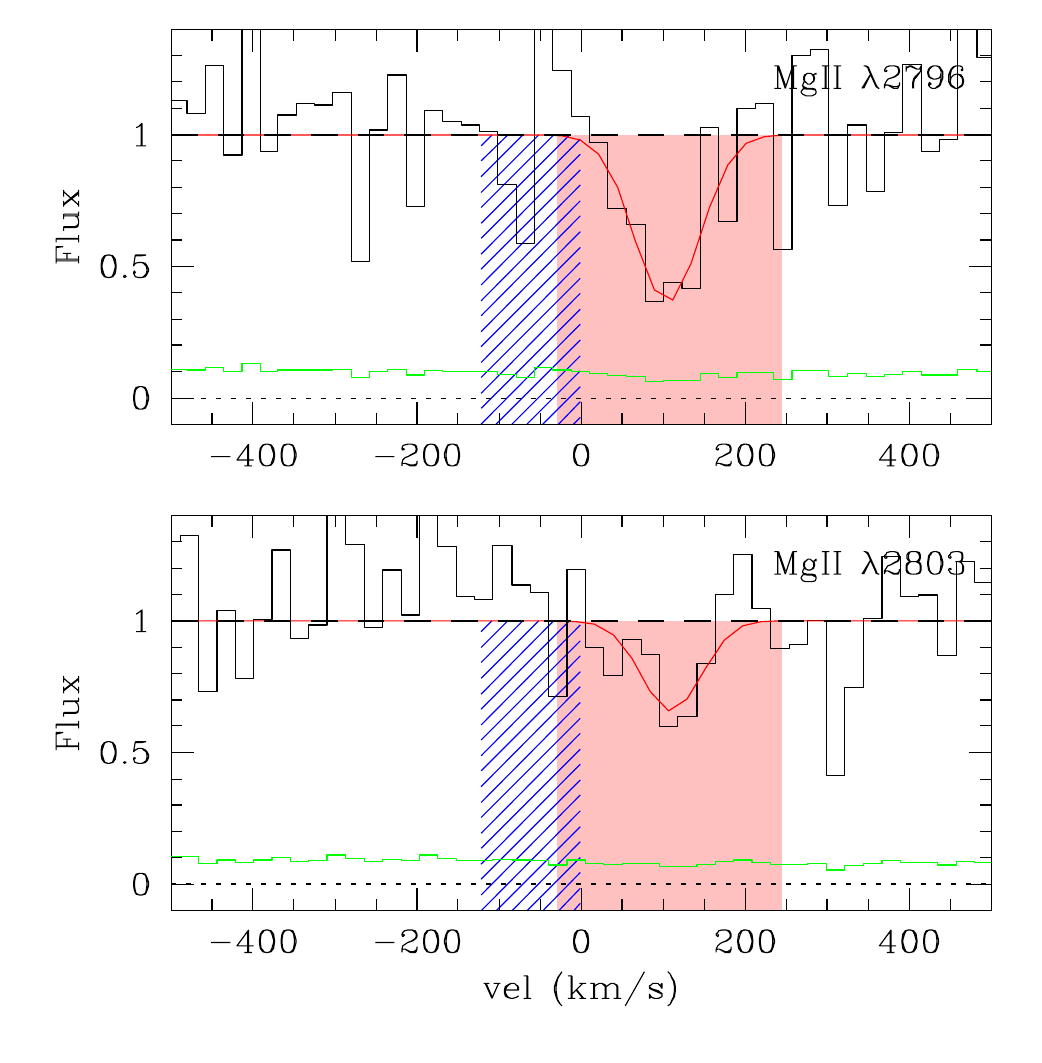}
}
\caption{{\bf (a)} {\it Top left}: Kinematic model of conical wind for J081420G1
viewed on the sky plane   where the ${ x}$-axis corresponds to the major axis and the ${ y}$-axis with the minor axis.  
 The solid oval represents an inclined disk and the black circles represent 
the conical outflow.
{\it Top right}: Side view where the ${ z}$ coordinate corresponds to the line-of-sight  direction, with the observer to the left.
{\it Bottom left}: Map of the averaged line-of-sight  velocities  (\kms) of the clouds.
{\it Bottom right}: Distribution (convolved with the LRIS resolution)
of the cloud $ z$-velocities at the observed QSO location.
The normalization of this distribution is inversely proportional to the impact parameter $b$
(see Eq.~\ref{eq:Ntrans}).
{\bf (b)} Observed \MgII\ kinematics from the LRIS spectra taken by \citet{KacprzakG_11a}, where 0~\kms\
corresponds to the galaxy systemic redshift. 
The solid line (red) shows a multi-Gaussian fit to the absorption features
to help the reader identify the \MgII\ absorption profiles.
The pink (grey shaded) and blue (grey hatched) areas show the range of velocities accounted by our wind model and by the halo-disk model of \citet{KacprzakG_11a}, respectively.
}
\label{fig:J0814}
\end{figure*}

Figure~\ref{fig:J0814}(a) shows an example of the wind model  
for the galaxy J081420G1 (towards the quasar SDSSJ081420.19$+$383408.3)
whose inclination is   $i\sim35^{\circ}$, and azimuth angle is $|\alpha|\sim80^{\circ}$. 
The top left panel shows the cone view face-on  and
the top right panel shows a side view of the cone.
The solid-blue oval represents the inclined disk and the black circles represent 
the conical outflow.
The bottom left shows the average  ${z}$-velocities 
of the  clouds as a function of position. 
The QSO location is represented as the filled circle.
 The bottom right panel shows the line-of-sight velocity
distribution of the clouds  at the location of the quasar.
The distribution is convolved with an instrumental resolution of $\sim150$~\kms, corresponding to
the LRIS data of \citet{KacprzakG_11a}.
The LRIS spectra is shown in Figure~\ref{fig:J0814}(b).
The pink (grey shaded) and blue (grey hatched) areas show the range of velocities accounted by our wind model and by the halo-disk model of \citet{KacprzakG_11a}, respectively.

In this particular case, we choose the model where the cone is pointing away from the observer  since the observed \MgII\ absorption is redshifted with respect to the galaxy systemic velocity (0~\kms).
The  wind speed $V_{\rm out}$ is tuned to match the observed
velocity range. For this particular case, we find that  $V_{\rm out}\sim200\pm25$~\kms\
produces a good match to the data.

\begin{figure*}
\subfigure[]{\includegraphics[width=11cm]{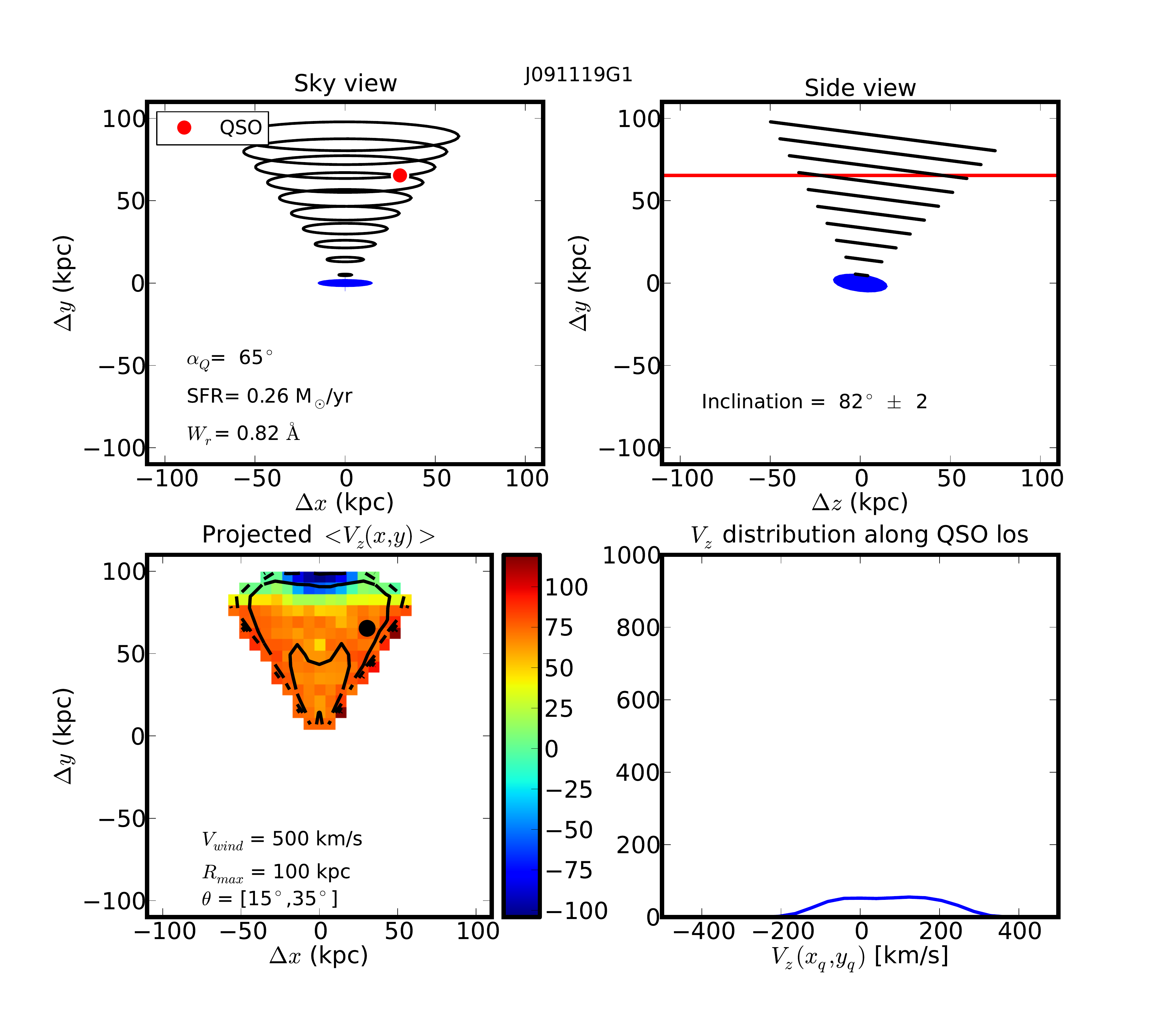}
}
\subfigure[]{
\includegraphics[width=5.5cm]{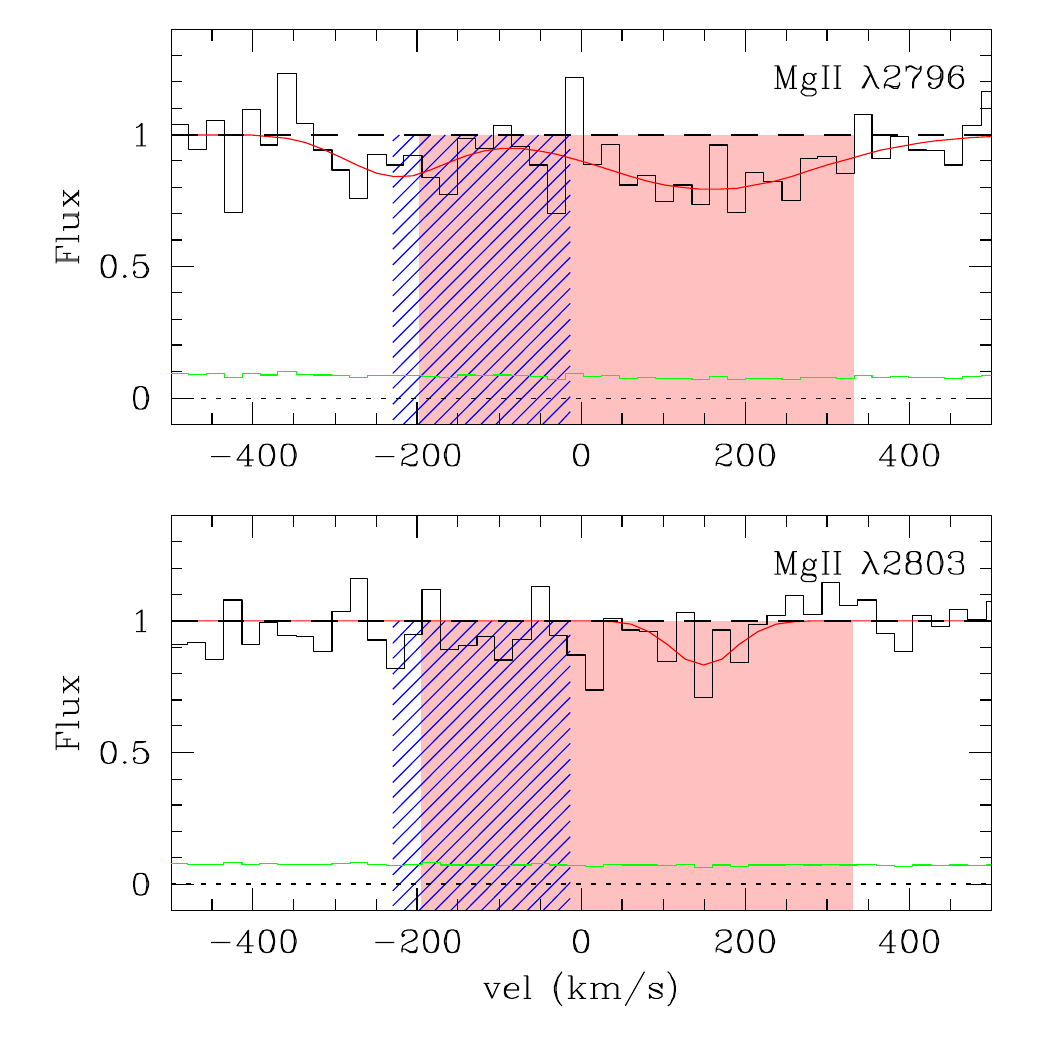}
}
\caption{{\bf a)}: Kinematic model  of conical wind for J091119G1 as in Fig.~\ref{fig:J0814}.
{\bf b)}: the observed \MgII\ kinematics with respect to the systemic velocity  as in Fig.~\ref{fig:J0814}. }
\label{fig:J0911}
\end{figure*}

Figure~\ref{fig:J0911} shows the wind model for J091119G1 towards
the quasar SDSSJ091119.16$+$031152.9.  This sight-line has one with the highest impact parameter ($b=71$~kpc). 
 While the equivalent width is relatively small ($\EW\sim0.8$~\AA),
the absorption is spread over a very wide range of velocities (from $-300$ to 300~\kms).  In addition, the galaxy is seen almost perfectly edge-on ($i\sim80^{\circ}$). As a result, our constant $V_{\rm out}$ model requires
a large wind speed of $\sim500$~\kms. We also note that the \MgII\ profile shows little or no absorption around $V_{\rm sys}$, which indicates either the absorbing material traces the edges of the cone or that this sight-line caught the wind as it stalls, in which case our assumption of pure radial wind velocities break down.

\begin{figure*}
\subfigure[]{
\includegraphics[width=11cm]{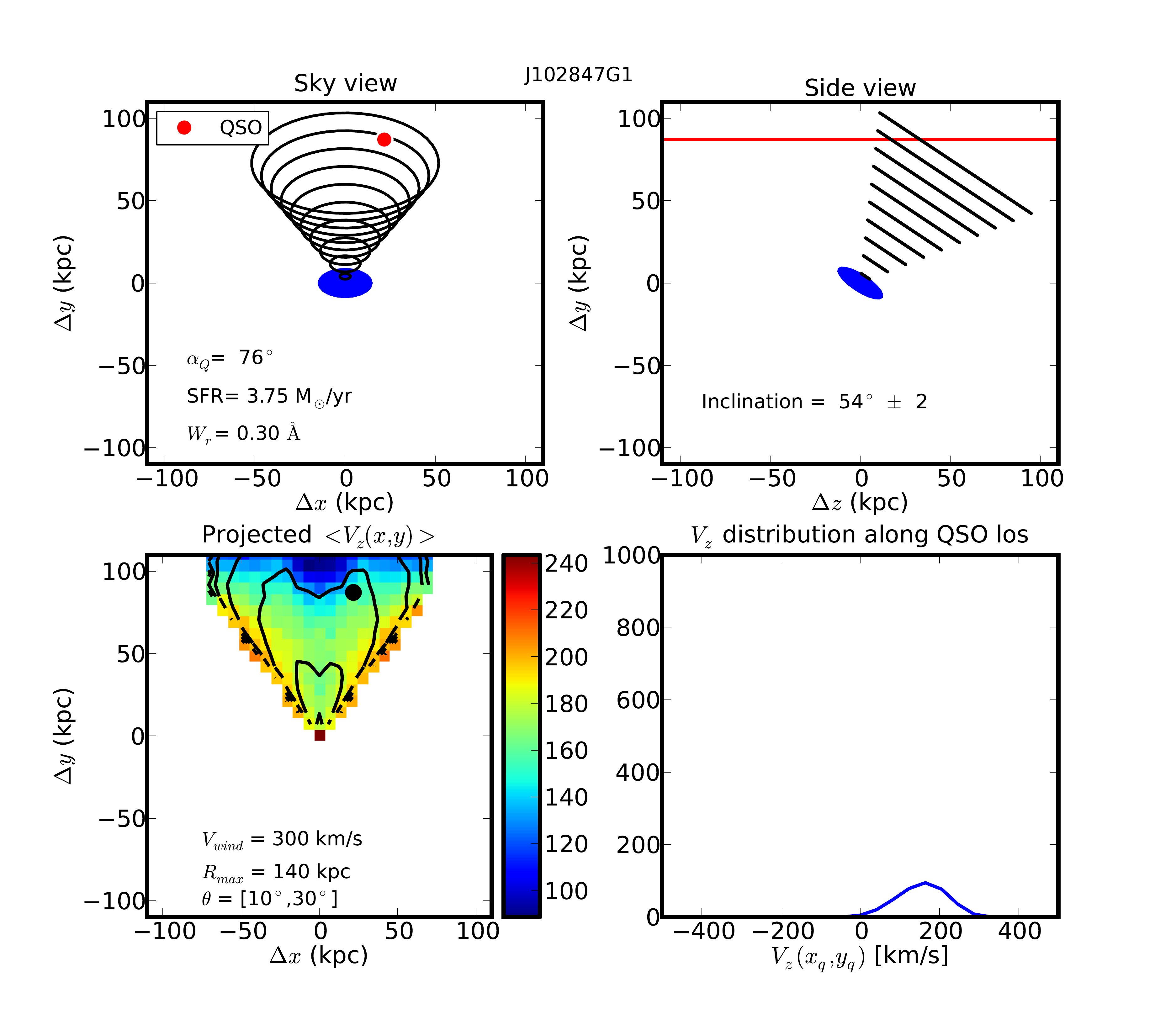} 
}
\subfigure[]{
\includegraphics[width=5.5cm]{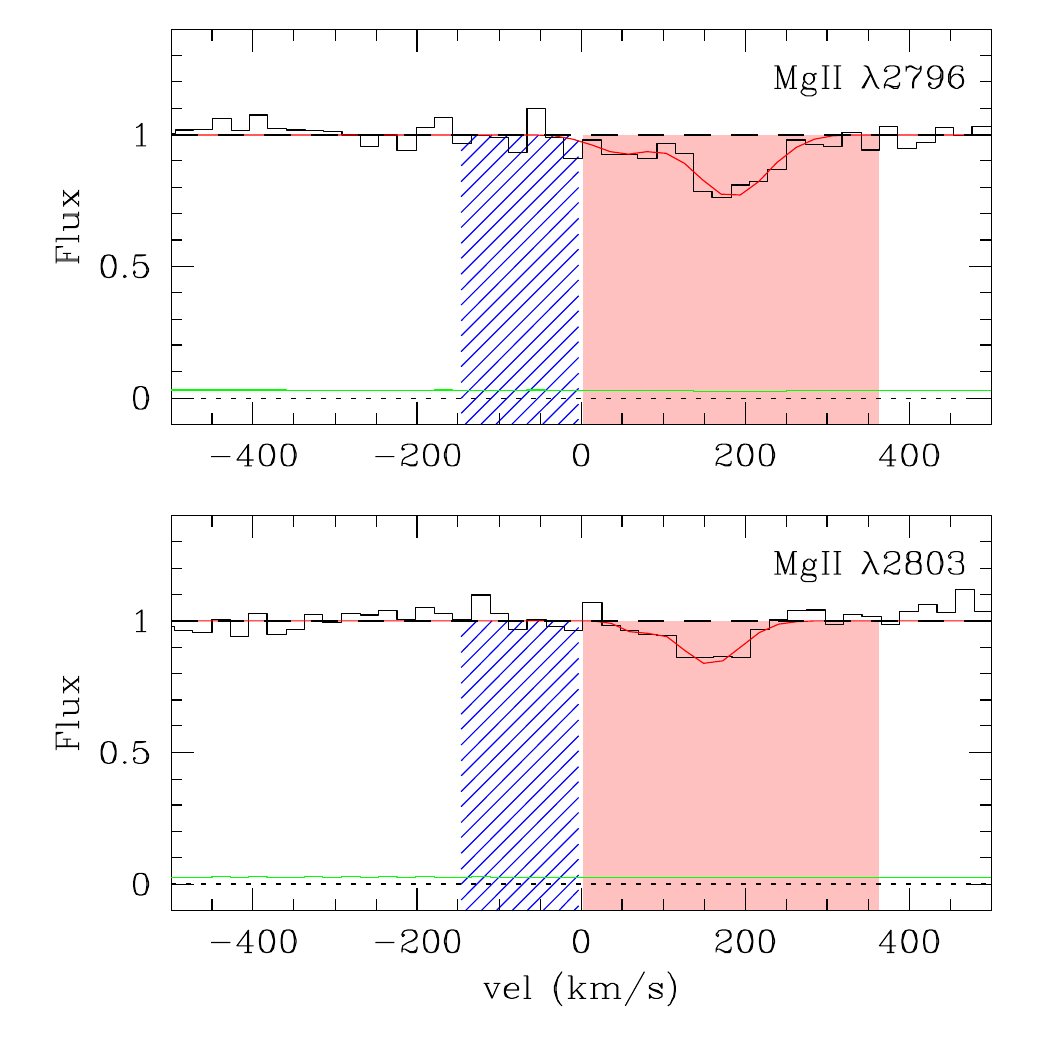}
}
\caption{{\bf a)}: Kinematic model  of conical wind for J102847G1 as in Fig.~\ref{fig:J0814}.
{\bf b)}: the observed \MgII\ kinematics with respect to the systemic velocity  as in Fig.~\ref{fig:J0814}. }
\label{fig:J1028}
\end{figure*}

Figure~\ref{fig:J1028} shows the wind model for J102819G1 towards
the quasar SDSSJ102847.00$+$391800.4.  The main difference with the previous example is that this galaxy is less inclined with $i\sim55^{\circ}$. As a result,
the wind produces absorption only redward of the galaxy systemic velocity. Because 
of the large impact parameter  ($b=89$~\kpc) of the quasar, we extended our wind
model to $R_{\rm max}=140$~\kpc\ and the inferred radial wind speed is $V_{\rm out}\simeq300$~\kms.
Note that the observed velocity profile is rather well reproduced.

\begin{figure*}
\subfigure[]{
\includegraphics[width=11cm]{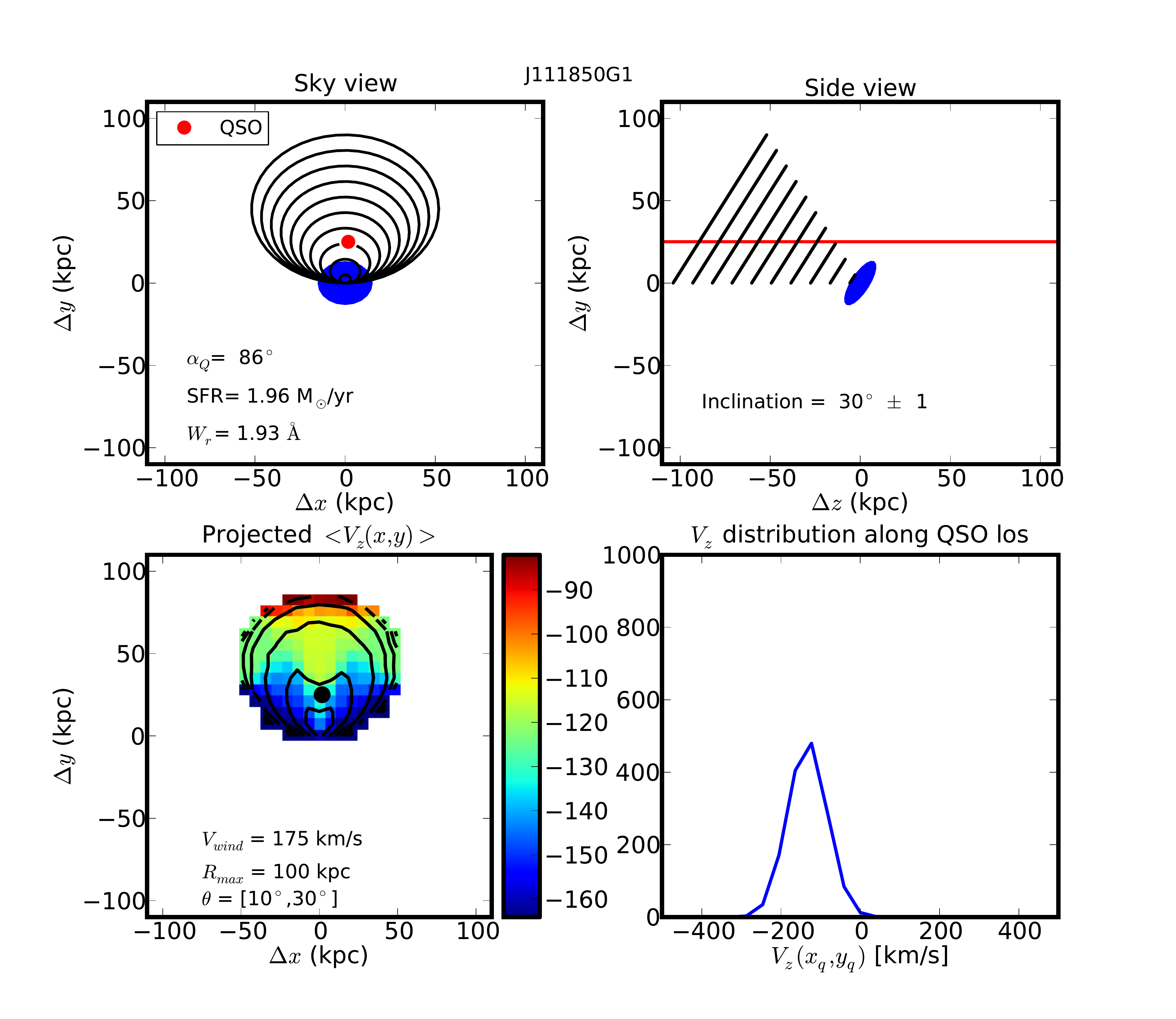}	
}
\subfigure[]{
\includegraphics[width=5.5cm]{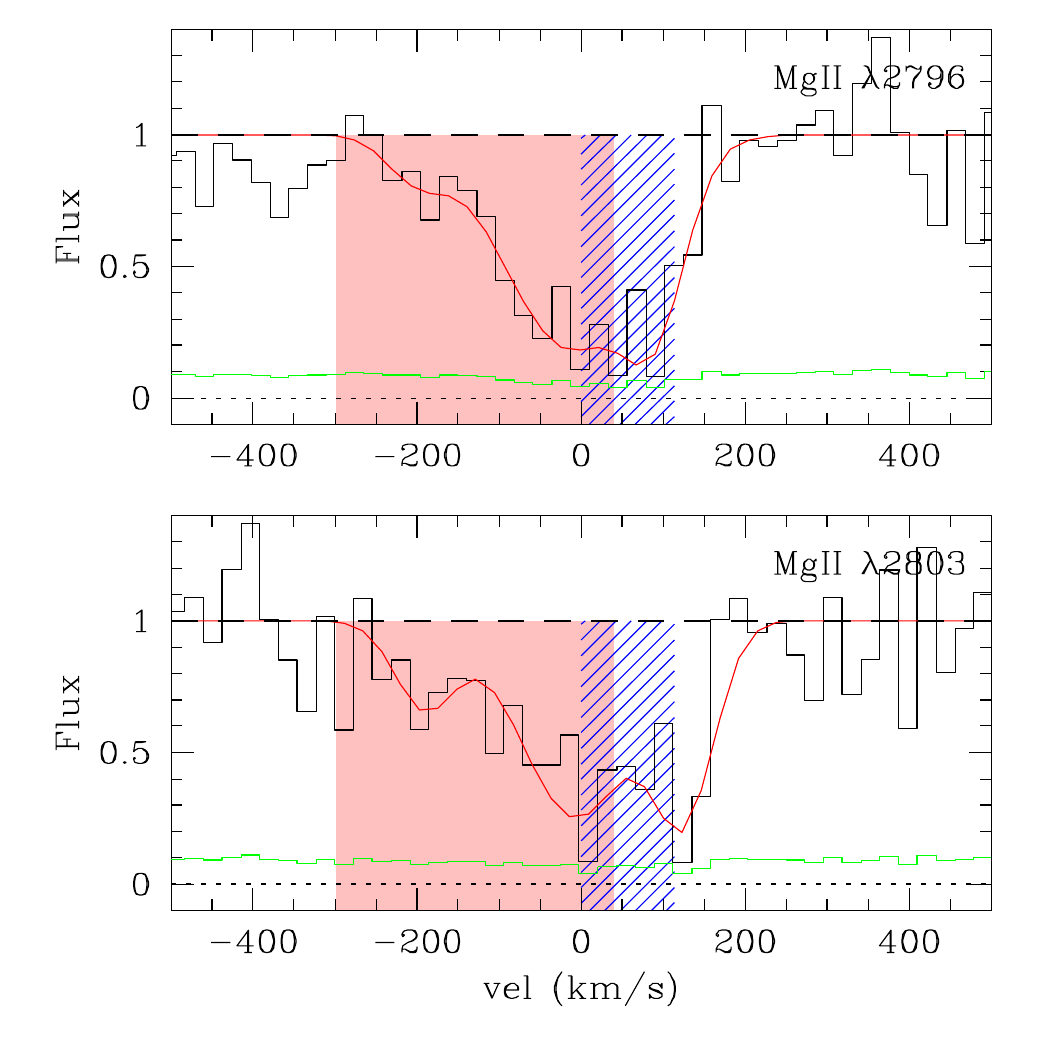}
}
\caption{{\bf a)}: Kinematic model  of conical wind for J111850G1 as in Fig.~\ref{fig:J0814}.
{\bf b)}: the observed \MgII\ kinematics with respect to the systemic velocity  as in Fig.~\ref{fig:J0814}.    Given the low inclination,
this sight-line is likely contaminated by absorption from the disk.
Indeed,  the blue hatched area shows the disk-halo model of \citet{KacprzakG_11a}. }
\label{fig:J1118}
\end{figure*}

Figure~\ref{fig:J1118} shows the wind model for J111850G1 towards
SDSSJ114518.47$+$451601.4.  This galaxy has SFR$=3.75$~\mpy, is less inclined with $i\sim30^{\circ}$,
and the quasar impact parameter is $b\simeq25$~kpc. As a result, this sight-line
likely intercepts the wind and the disk at the same time, resulting
in significant absorption at $V_{\rm sys}$. 
As a result, our model, which does not include a disk component, 
cannot reproduce the entire \MgII\ profile and the wind speed is less constrained.
In fact, the disk model of \citet{KacprzakG_11a} can reproduce the kinematics not accounted
by our wind model (blue hatched).

\begin{figure*}
\subfigure[]{
\includegraphics[width=11cm]{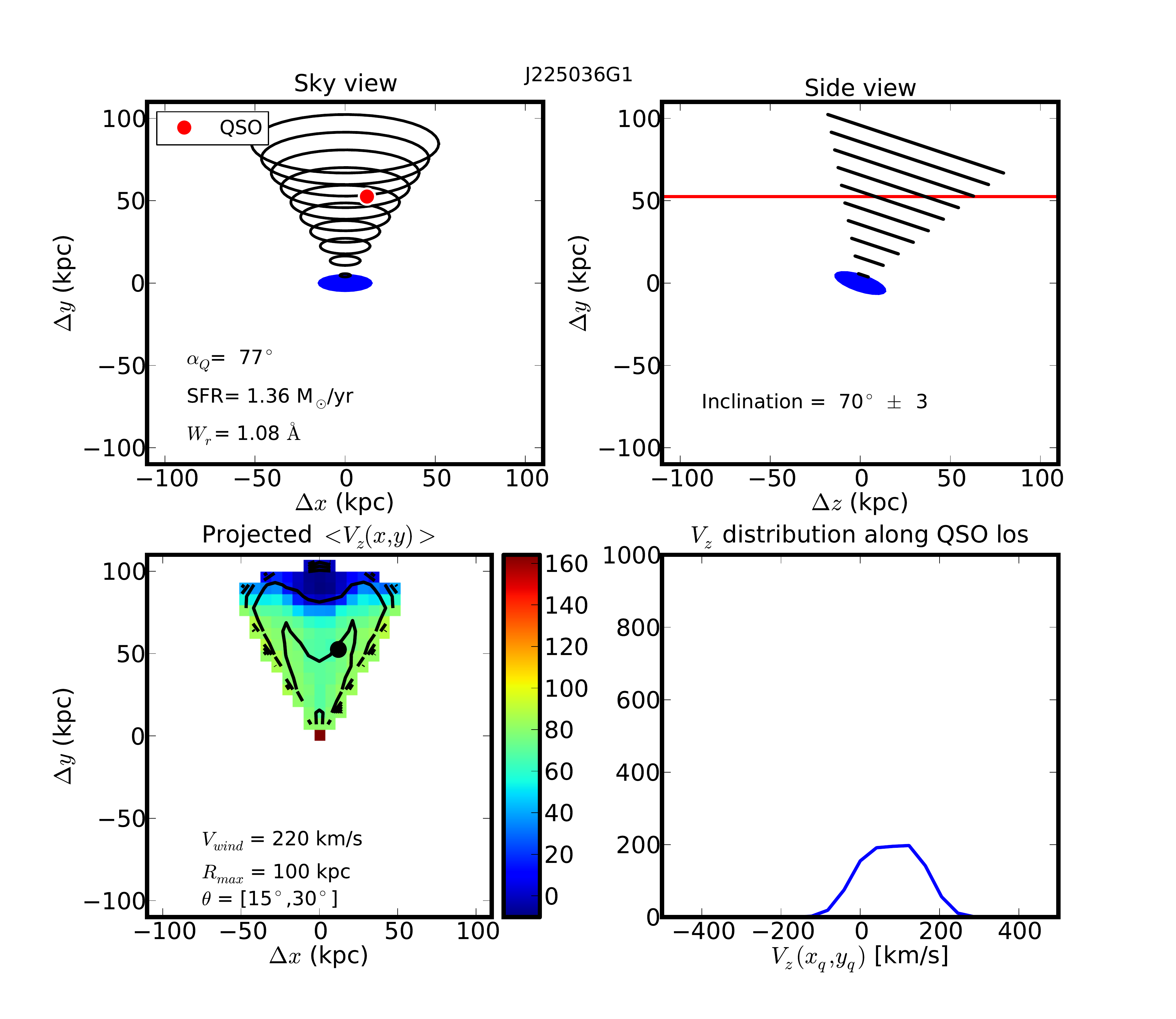}	
}
\subfigure[]{
\includegraphics[width=5.5cm]{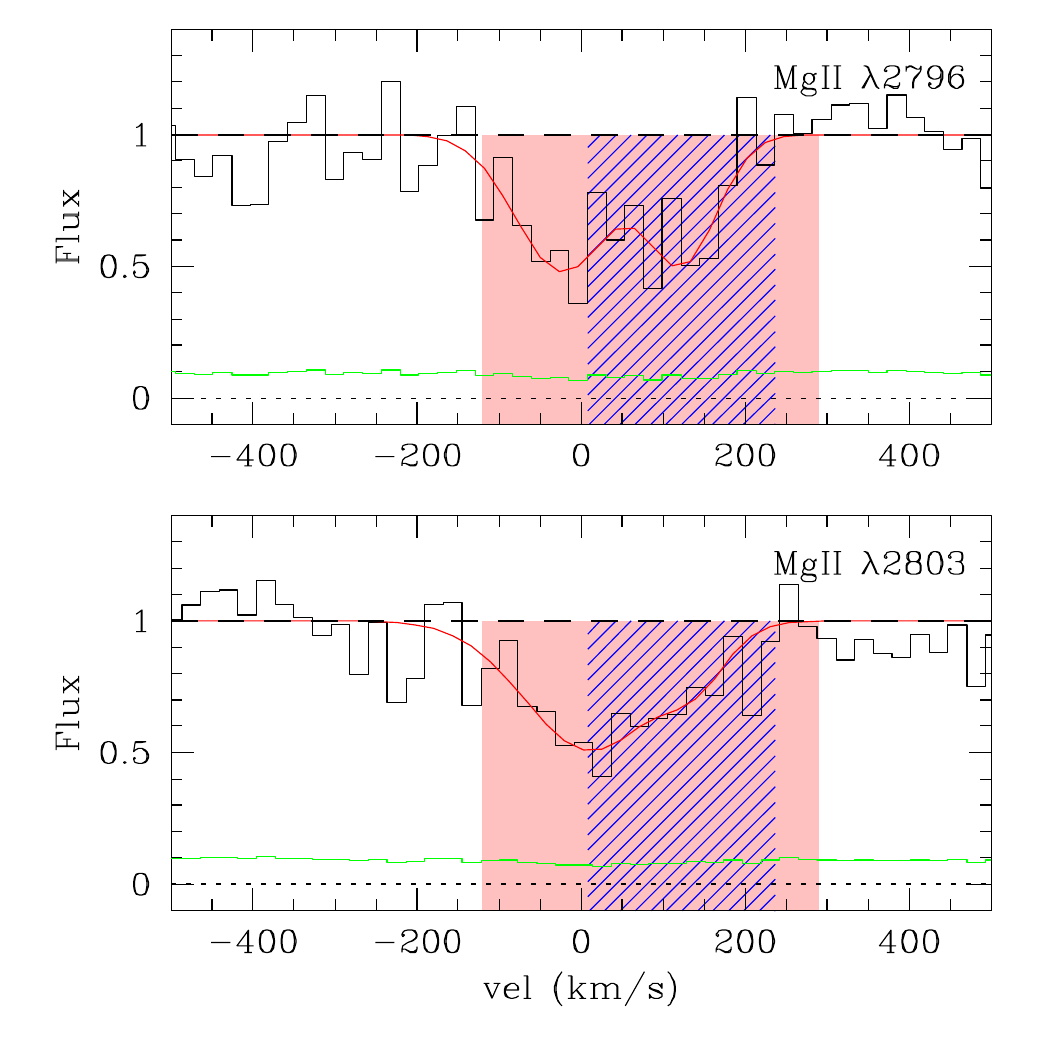}
}
\caption{{\bf a)}: Kinematic model  of conical wind for J225036G1G1 as in Fig.~\ref{fig:J0814}.
{\bf b)}: the observed \MgII\ kinematics with respect to the systemic velocity  as in Fig.~\ref{fig:J0814}. }
\label{fig:J2250}
\end{figure*}

Figure~\ref{fig:J2250} shows the wind model for J225036G1 towards
SDSSJ225036.72$+$000759.4. This galaxy has a moderate SFR of 1.36~\mpy, is highly inclined with $i\sim70^{\circ}$ and the quasar impact parameter is
$b\sim50$~\kpc.  The radial wind speed  $V_{\rm out}$ is inferred to be $\sim200$~\kms.
Given this geometric configuration with the inclination approaching 90$^{\circ}$,
 the sight-line probes a wide range of velocities, a feature that is very consistent with the observed \MgII\ kinematics.  Note also that the profile suggests that the cone is hollow, i.e. that the \MgII\ absorption
arise on the edges on the cone, which is the reason why we used $\theta_{\min}=15^{\circ}$ in this case.

\begin{figure*}
\subfigure[]{
\includegraphics[width=11cm]{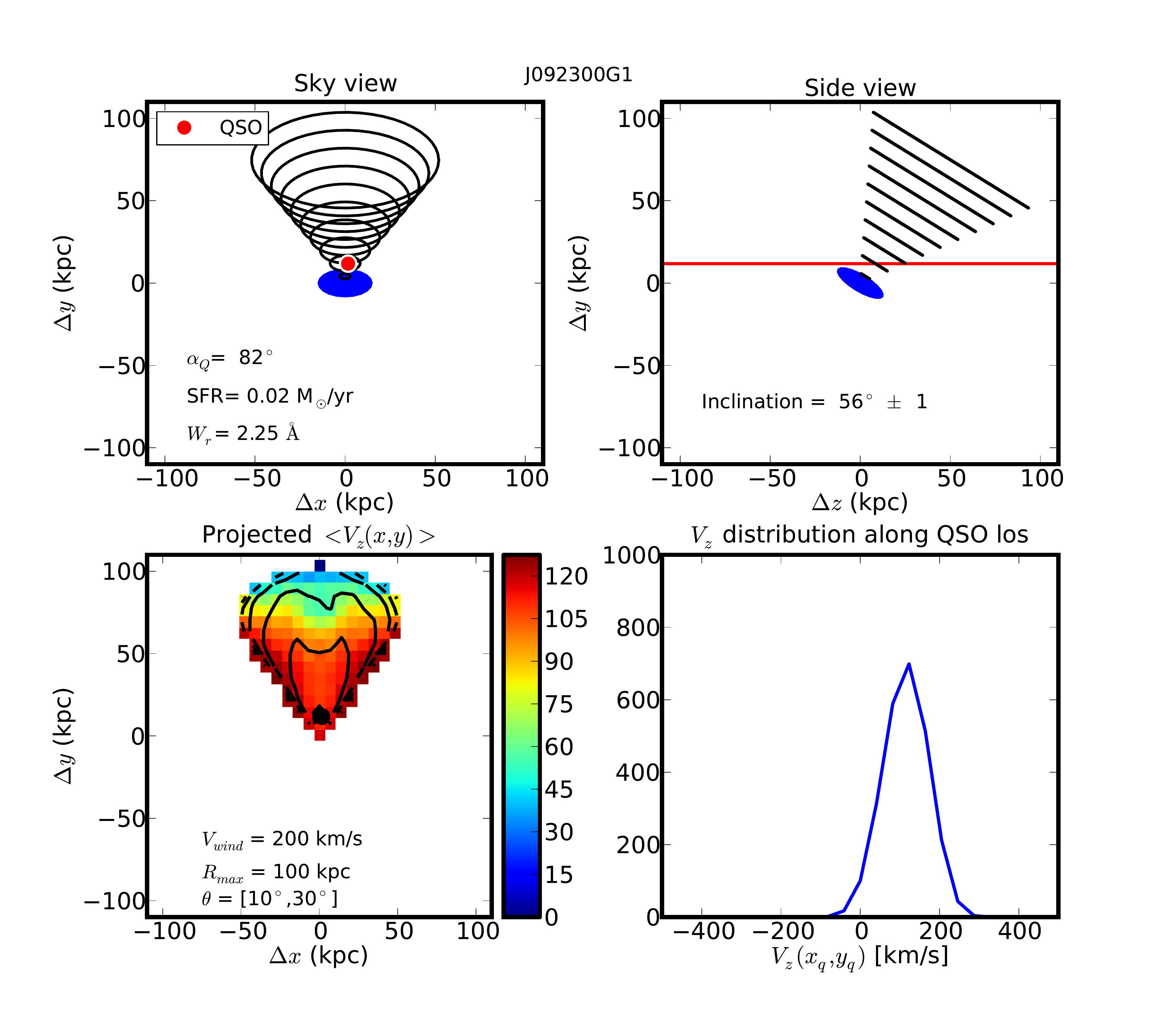}	
}
\subfigure[]{
\includegraphics[width=5.5cm]{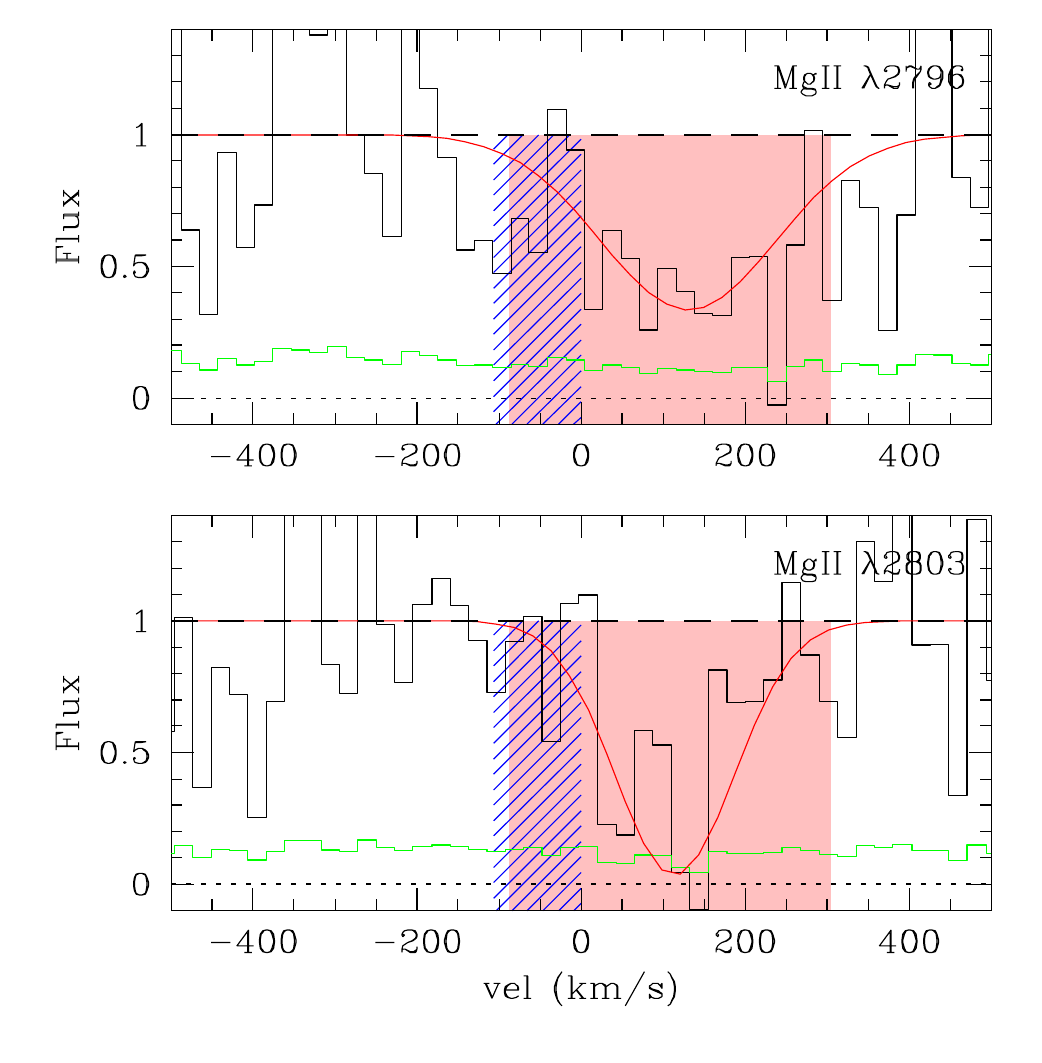}
}
\caption{{\bf a)}: Kinematic model  of conical wind for J092300G1 as in Fig.~\ref{fig:J0814}.
{\bf b)}: the observed \MgII\ kinematics with respect to the systemic velocity  as in Fig.~\ref{fig:J0814}. }  
\label{fig:J0923}
\end{figure*}

We next discuss two pairs (J092300G1, J114518G1) 
whose classification based on $\alpha$ was misleading
once the global aspects of the geometry are taken into account.  
In particular, the sight-line towards
SDSSJ092300.67$+$075108.2  has the lowest
impact parameter ($b=12$~kpc).
Figure~\ref{fig:J0923} reveals that
the low impact parameter combined with the high galaxy inclination implies that the 
sight-line can intercept the other parts of the galaxy in spite of having
a high azimuth angle  $|\alpha|=82^{\circ}$.  
Hence, this galaxy-quasar pair is classified as 'ambiguous'.
Furthermore, the galaxy has a low SFR of 0.02~\mpy, red colors and an early type morphology (S\'ersic index of $n\sim5$).

\begin{figure*}
\subfigure[]{
\includegraphics[width=11cm]{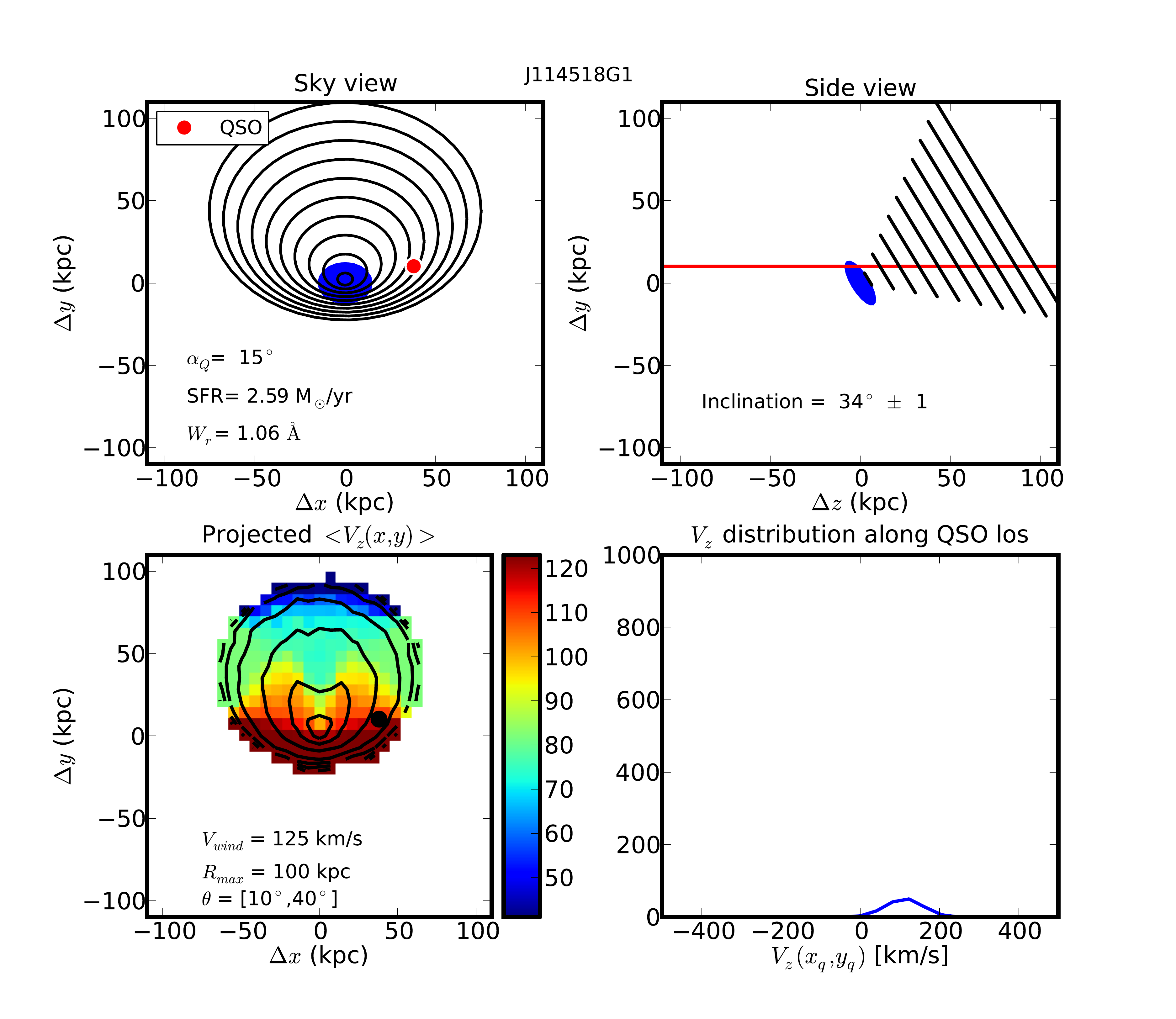} 
}
\subfigure[]{
\includegraphics[width=5.5cm]{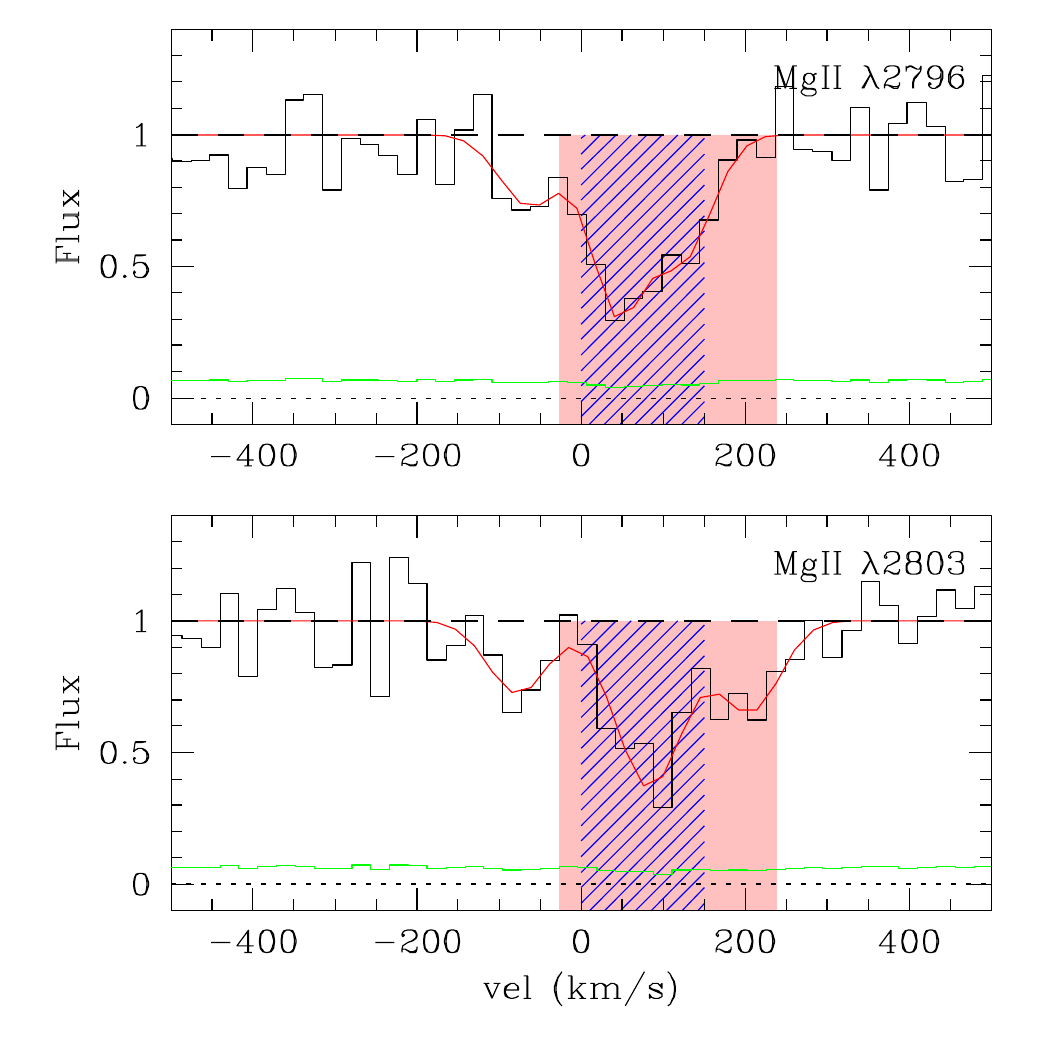}
}
\caption{Kinematic model (a) of conical wind for J114518G1 compared the observed \MgII\ kinematics (b) with respect to the systemic velocity  as in Fig.~\ref{fig:J0814}.  }
\label{fig:J1145}
\end{figure*}

The other ambiguous case is  J114518G1 towards
SDSSJ114518.47$+$451601.4.  This sight-line has a low $\alpha$ of $\sim15^{\circ}$.
However,  the wind modeling (Figure~\ref{fig:J1145}) 
 reveals that the sight-line can very well intercept the wind given
the low galaxy inclination ($i\sim30^{\circ}$) if the cone extends a bit beyond our canonical 30~deg to 40~deg. The wind model can account for the \MgII\ kinematics   with an outflow speed of $V_{\rm out}\simeq125$~\kms.
In addition, the galaxy SFR is high with SFR$=2.59$\mpy, blue colors and an exponential profile,
which are additional reasons 
to consider this pair as a `wind' pair.

\subsubsection{Summary of the Wind Modeling}

In summary, we performed simple kinematic models for the  quasar-galaxy pairs  whose azimuth angles $|\alpha|$  are close to $90^{\circ}$.  Figures~\ref{fig:J0814}--\ref{fig:J2250} show  that  a simple geometric model can account surprisingly well for the observed \MgII\ kinematics of four out of five  QSO-galaxy pairs classified as `windy'  and for two other pairs initially classified as `disky' based on their $\alpha$ measurement  (see table~\ref{table:measurements}). 
 Conversely, the wind model
is unable to account for the kinematics of the sight-lines with $\alpha\sim0^{\circ}$  since 
they do not intercept the conical flow.

In addition, the disk-halo model of \citet{KacprzakG_11a}, shown as the hatched
areas in Figures~\ref{fig:J0814}--\ref{fig:J2250}, has serious difficulties in reproducing the \MgII\  kinematics for the 'wind' sub-sample since the projected disk velocities are approximately 0~\kms\ along the minor axis. However,
in the case of J111850G1,
our wind model cannot reproduce the entire \MgII\ profile but the disk model of \citet{KacprzakG_11a} can account for the extra component.

Quantitatively, our analysis reveals
wind speeds $V_{\rm out}$ that are typically 100--300~\kms\ (listed in table~\ref{table:measurements}), i.e.  are on the order of the maximum circular speed $V_{\rm max}$ determined from the rotation curves by \citet{KacprzakG_11a}.
Based on these results, we will investigate the wind properties in more detail in the next section \S~\ref{section:winds}.  We discuss the properties of the 'disk' sub-sample in Appendix~\ref{appendix:disk}.

\subsection{Radial Dependence}
\label{section:radial}

\begin{figure}
\centering
\includegraphics[width=9cm]{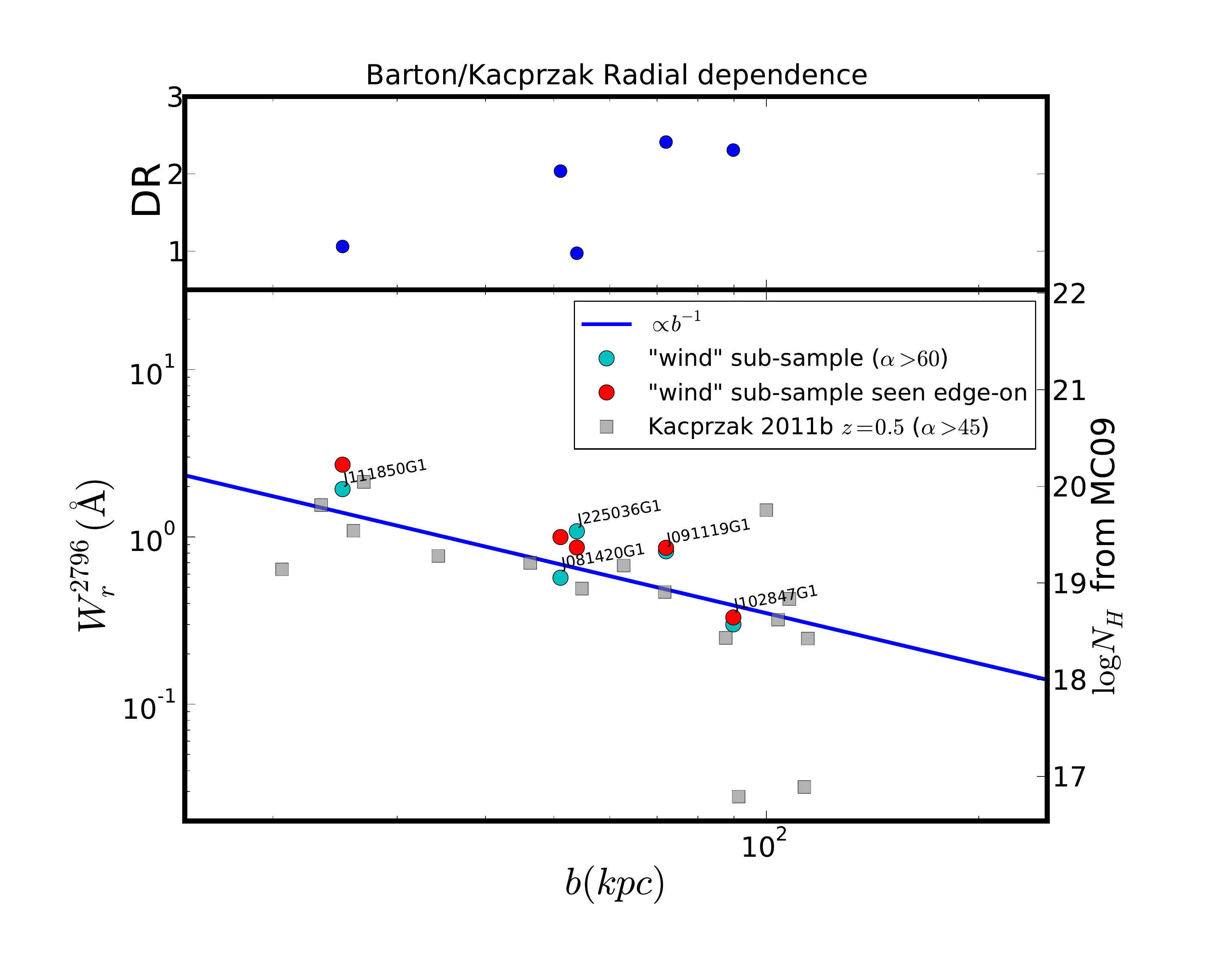}
\caption{\EW\ as a function of impact parameter $b$ for QSO-galaxy pairs 
classified as `wind', i.e. with $|\alpha|>60$.
The cyan   circles represent the observed  \EW, and the red circles show the \EW\ corrected
for inclination effects (see text), for the edge-on case where $i=90^{\circ}$. 
The squares show the $z\sim0.5$ QSO-galaxy pairs of \citet{KacprzakG_11a} for a similar sub-sample with the azimuth angle $\alpha>45$. 
The expected radial dependence ($\propto b^{-1}$) for sight-lines intercepting a cone at 90$^{\circ}$ is shown by the solid line
and is a good description of these two data sets. 
The right $y$-axis shows the corresponding \NHI\ using the \EW-\NHI\ relation
from \citet{MenardB_09a} (MC09). The radial dependence of \NHI\ is supported
by the \MgII\ doublet ratio  shown in the top panel.}
\label{fig:radial}
\end{figure}

Using the sub-sample of $z\sim0.1$ QSO-galaxy pairs that we associate with galactic outflows,
we investigate the radial dependence of the \MgII\ absorption.
Figure~\ref{fig:radial} shows the observed \EW\ (cyan circles) as a function of impact parameter $b$.  Because the \EW\ can be skewed due to the various inclination effects,
we corrected the \EW\ (red circle) to a common inclination of $i=90^{\circ}$. 
The corrections are 
 calculated  by comparing the total number of clouds intercepted at the observed $i$ to 
that number for $i=90^{\circ}$. The most significant correction
   is for J081420G1 (see Fig.~\ref{fig:J0814})
because that  sight-line intercepts a small fraction of the cone compared
to  the  edge-on situation.

For comparison, we show in Figure~\ref{fig:radial} the radial dependence of the cool halo gas 
from  the $z\sim0.5$ sample of QSO-galaxy pairs    collected and analyzed in 
\citet{KacprzakG_07a,KacprzakG_11b} (squares). Since, the $|\alpha|$ distribution is also bi-modal for
this sample \citep{ChurchillC_12a,BordoloiR_12a}, we only show those with $|\alpha|>45^{\circ}$
and  whose  uncertainty in $\alpha$ is less than 30$^{\circ}$ (3$\sigma$) in order to remove ambiguous cases. 
Figure~\ref{fig:radial} shows that the $z=0.1$ and $z=0.5$ data sets share a common \EW--$b$ relation which goes approximately as $b^{-1}$.
A formal linear fit to the data gives $b=-1.1\pm0.5$.

Our radial dependence is significantly different that the one determined by \citet{BordoloiR_11a} around of $z\sim1$ inclined disks using background galaxies. However, we refrain from any direct comparison
since the background galaxy technique used by  \citet{BordoloiR_11a} does not give a measure of the collective absorption of \MgII\  `clouds', but  gives a measure  of the radial dependence  of the covering fraction $C_f(r)$~\footnote{At $z\sim2$, the covering fraction
was found to go as $C_f(b)\propto b^{-0.4\pm0.2}$ by \citet{SteidelC_10a},
 close to the theoretical expectation of $C_f\propto b^{-2/3}$
for adiabatically  expanding clouds moving in a hot medium \citep{MartinC_09a}.} 
because the background sources are naturally extended~\footnote{We note the interpretation  of the radial dependence in \citet{BordoloiR_11a} using isothermal profiles from \citet{TinkerJ_08a} is not appropriate here since the model of \citet{TinkerJ_08a} was tuned to quasar absorbers.}.

Given that there is an empirical relation between \EW\ and the total column density \NHI\ \citep{MenardB_09a}, we show on the right $y$-axis of Figure~\ref{fig:radial}, the
corresponding column density $N$. 
 The expected radial dependence of the column density $N(b)$  
 for an optically thin medium, whose density $\rho(r)$ is  geometrically diluted
   $\left [\rho(r)=\rho_0\left({r_0}/{r}\right)^2\right ]$,  is 
\begin{eqnarray}
N(b)&\propto& \int \frac{{\rm d}x}{b^2+x^2} \propto\frac{1}{b} \label{eq:radial}
\end{eqnarray}
where the integral is performed perpendicularly to the cone
(Eq.~\ref{eq:Ntrans}). 
Eq.~\ref{eq:radial} shows that $N(b)$ is expected to go as $\propto b^{-1}$.
The solid line in Figure~\ref{fig:radial} shows that this is very good description of the data. The scatter around the solid is only 0.20~dex.
The match between the data and the expected radial dependence may seem surprising given that Eq.~\ref{eq:radial}
assumes an optically thin medium whereas the \EW\ absorption can be optically thick. 
This can be understood if one realizes that the \EW\ is also proportional
to the number of sub-components or clouds \citep[e.g.][]{BergeronJ_91a,ChurchillC_03a,CheloucheD_07a}.

In summary, Figure~\ref{fig:radial} shows that, for QSO-galaxy pairs associated with galactic outflows, there is a tight correlation between \EW\ and impact parameter $b$, 
following the expected $b^{-1}$ dependence. 
The scatter around $b^{-1}$ is very small, only 0.20~dex for this sub-sample. 
While this anti-correlation has been known for two decades
\citep[e.g.][]{LanzettaK_90a,SteidelC_95b,BoucheN_06c,ChenHW_08a}, 
the scatter has been previously shown to be 0.5~dex \citep{ChenHW_10a,ChurchillC_12a}.  

\citet{KacprzakG_11b} argues that this scatter is a function of the host inclination,
and \citet{ChenHW_10a} argues that this scatter is 
 correlated with stellar mass and perhaps also with SFR \citep{ChenHW_10b}.  
In Appendix~\ref{appendix:disk}, we discuss the \EW-$b$ relation for the 'disk' sub-sample and show that the large  scatter
is caused by the mixing of several physical mechanisms in \MgII\ samples, namely extended gaseous disks and galactic winds \citep[see also][]{ChurchillC_12a,BordoloiR_12a}.
It remains to be demonstrated whether the stellar mass or SFR
dependence applies to the `disk' or `wind' sub-sample.

\section{Extracting Wind Properties}
\label{section:winds}

Having established that the \MgII\ kinematics in QSO-galaxy pairs with $|\alpha|\sim90^{\circ}$ are consistent with intercepting entrained material in galactic winds, in this section,
we focus on the terminal velocity (\S~\ref{section:Vwind}) and the mass outflow rate (\S~\ref{section:Mout}) of these galactic winds.

\subsection{Terminal velocity}
\label{section:Vwind}

The cool gas in galactic winds traced by \MgII\ 
may be driven either by the kinetic energy
of supernova ejecta from the entrainment in the hot wind
\citep[e.g.][]{ChevalierR_85a,HeckmanT_90a,StricklandD_00a},
   by  momentum injection from the radiation
pressure \citep{MurrayN_05a,MurrayN_11a,SharmaM_11a} or by cosmic ray pressure
\citep[][and references therein]{EverettJ_08a}. 
Because these predict different scalings for the wind velocity $V_{\rm out}$
with galaxy mass and SFR, it is important to investigate whether $V_{\rm out}$
varies with other galaxy properties.

In our sample of galaxies which have SFRs of a few \mpy, 
we find that the outflow speeds $V_{\rm out}$   are typically 100--300~\kms\ (listed in Table~\ref{table:measurements}) using the modeling presented in  section~\ref{section:model}.
These relatively low speeds are  of the order of the circular velocity $V_{\rm max}$. 
We note that  $V_{\rm out}$ appears to increase with impact parameter. 
We caution that a  larger sample is required in order to put strong constraints on $V_{\rm out}(b)$,
a function that is directly related to  the acceleration mechanism in the wind \citep[e.g.][]{MurrayN_05a,MurrayN_11a}.

We find no   correlation between $V_{\rm out}$ and SFR. 
This could either be due  (i) to our small range in SFRs or  (ii) to
our \Ha-derived SFRs which may not be related to the SFRs that occurred
when the material was launched given that the travel time to the observed impact parameter can be significant. Typically, the travel time is of the order of 0.5~Gyr (see Table~\ref{table:measurements}).

Our measurements of outflow speeds allow us to address the following question:
Are these velocities sufficient to expel the gas from the
galaxy into the IGM or will the gas eventually fall back onto the galactic disk?
The escape velocity $v_{\rm esc}$ for an isothermal sphere is 
\begin{equation}
V_{\rm esc}= V_{\rm max}\;\sqrt{2[1+\ln (R_{\rm vir}/r)]} \label{eq:vesc}
\end{equation}
where $V_{\rm max}$ is the maximum circular velocity (a proxy for $V_{\rm vir}$) and
$R_{\rm vir}$ is the virial radius.
Since our galaxies are $L^*$ galaxies with halo mass around $10^{12}$\msun, 
their virial radius is approximately $R_{\rm vir} \equiv {V_{\rm max}}/{10 H(z)}\simeq250$~kpc, where $H(z)$ is the Hubble constant at redshift $z$. 
Using Eq.~\ref{eq:vesc}, the escape velocity  $V_{\rm esc}$ is 2.5, 2.3, 1.8 $\times v_{\rm rot}$ 
at $b=$10, 50, 100 kpc, respectively. 
We find that most of our galaxies have wind speeds that are  about half the escape velocity. 

\begin{figure}
\centering
\includegraphics[width=8.5cm]{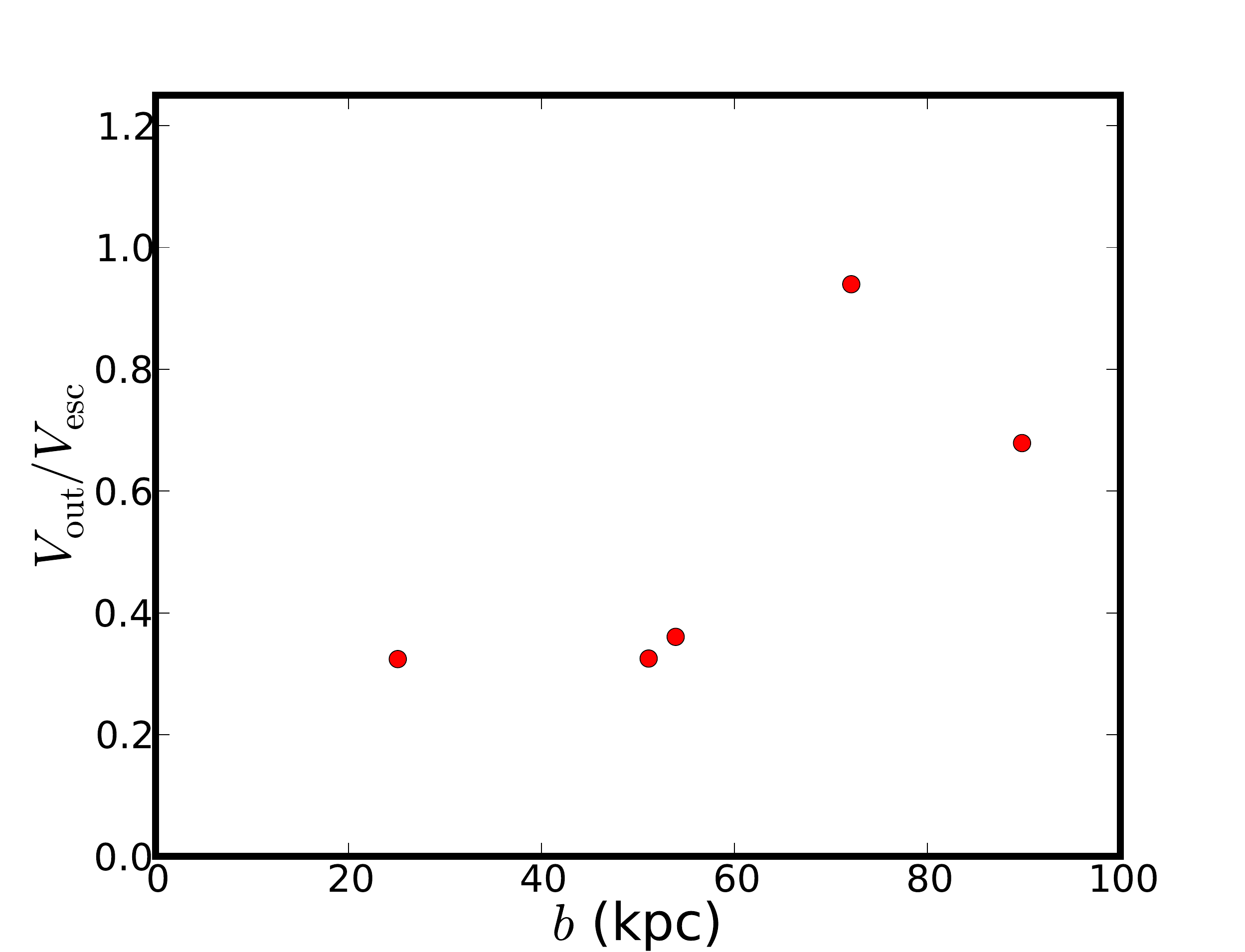}
\caption{Radial dependence of the outflow speed $V_{\rm out}$ with respect to the escape velocity $V_{\rm esc}(b)$. 
This shows that the cool material proved by \MgII\ is traveling at speeds
 $\la V_{\rm esc}$.}
\label{fig:vesc}
\end{figure}

Figure~\ref{fig:vesc} shows the outflow speeds relative to the escape velocity as a function of impact
parameter $b$ for our sample of 6 sight-lines.  
We use the individual rotation velocities $(V_{\rm max}$) from
\citet{KacprzakG_11a} and appropriate virial radii in Eq.~\ref{eq:vesc}.
This ratio $V_{\rm out}/V_{\rm esc}$ is $\la1$ showing that the cool material
probed by \MgII\ is traveling at speeds close to the escape velocity.
Interestingly,  the ratio $V_{\rm out}/V_{\rm esc}$ is about unity
for only two sight-lines, which
are the two  with the largest impact parameter (J102847G1, J0911119G1).

\subsection{Outflow rates}
\label{section:Mout}

Given that our $\alpha$ distribution combined with the results of \citet{BordoloiR_11a} and \citet{ChenY_10a}  clearly demonstrate
the presence of  bi-conical outflows within $\theta_{\rm max}\sim30^{\circ}$ of the minor axis of star-forming galaxies,
we can estimate the cold gas mass outflow rate corresponding to
 such a configuration. 
The outflow rate $\dot M_{\rm out}$ for a mass-conserved flow moving at a speed $V_{\rm out}$,
 with a total solid angle $\Omega_{\rm w}$,    is
 $\dot M_{\rm out}(r)\equiv \rho(r)\; \Omega_{\rm w} r^2\; V_{\rm out}$.
In the case of a radial sight-line looking `down-the-barrel',
the outflow rate reduces to
$\dot M_{\rm out}=\Omega_{\rm w} \overline N \;r_0 \; V_{\rm out}$ \citep[e.g.][]{HeckmanT_00a,MartinC_05a}
where the column is  
 $\overline N\equiv \int_{r_0}^{\infty} \rho(r) {\rm d}r = \rho(r_0)r_0$.
Similarly, in the case of transverse sight-lines at impact parameter $b$
  the outflow rate is:
\begin{eqnarray}
 \dot M_{\rm out}(b)&\simeq&  \frac{\pi}{2}\theta_{\rm max}\;N_{H}(b)\;b\;V_{\rm out} \label{eq:outflow}
\end{eqnarray}
 as derived in Appendix~\ref{appendix:outflow} (Eq.~\ref{eq:Moutflow:transverse}).
Inserting the numerical values for $\theta_{\rm max}\simeq30^{\circ}$, we have 
\begin{equation}
\dot M_{\rm out}(b)=0.41\;\mpy\frac{\mu}{1.5}\frac{\theta_{\rm max}}{30^{\circ}}
\frac{N_H(b)}{10^{19}\hbox{cm$^2$}}\frac{V_{\rm out}}{200~\kms}\frac{b}{25~\hbox{kpc}} \label{eq:outflow}
\end{equation}
where $\mu$ is the mean atomic weight.

The only unknown parameter in Eq.~\ref{eq:outflow} is the total gas column \NHI,
since the impact parameter $b$ and the wind speeds $V_{\rm out}$ are directly or indirectly constrained by the observations.
Fortunately, we can estimate
the mean \HI\ column using the empirical \MgII--\HI\ relation from \citet{MenardB_09a} \citep[see also][]{RaoS_06a,BoucheN_08a,RaoS_11a}  used already in Figure~\ref{fig:radial}.
Here, we make the implicit assumption that the relation holds for sight-lines associated with
galactic winds.  This assumption is supported by \citet{BoucheN_08a} who reported
that there are two populations of absorbers, one following the \HI-\EW\ relation 
(corresponding to outflows) with a high metallicity (half solar to solar), 
and another with roughly constant \HI\ with a low metallicity (1/30) corresponding to  typical DLAs/sub-DLAs. 
 A direct
determination of the gas column \HI\ would require  observations with the Cosmic Origin Spectrograph (COS).

The typical outflow rates derived from Eq.~\ref{eq:outflow} are $\sim$1--5~\mpy\ for all of our galaxies.
These are the most precise outflow rates derived for star-forming galaxies for the cold ($T\sim10^4$~K) gas~\footnote{Galactic winds are multi-phase phenomenon with potentially additional mass in the other phases not included here. However, most of the mass is likely contained in the cold phase while most of the energy in the hot phase.}.
Indeed, the uncertainties in $\dot M_{\rm out}$ are entirely dominated by the $\NHI$  uncertainties. Assuming 0.25~dex uncertainty for \NHI,
 the relative accuracy for the outflow rate is $\sigma(\dot M)/\dot M\simeq0.5$.
Compared to traditional spectroscopy, where both \NHI, $\Omega_w$ are only known to orders of magnitude
\citep[e.g.][]{HeckmanT_00a,PettiniM_02b}, our technique to estimate mass outflow rates is a leap forward. 

Given that the travel time of the low-ionization gas to the observed impact parameters is significant, a few 100 Myr (see Table~\ref{table:measurements}), we refrain from comparing our outflow rate to the instantaneous SFR. In  future work, we intend to compare them to the past SFR determined from stellar population analysis.
Overall, the cold gas mass outflow rate $\dot M_{\rm out}$ seems to be  $2$ times larger than the current SFR.

\section{Conclusions}
\label{section:conclusions}

In summary, we find that the azimuthal orientation of quasars with \MgII\ absorbers ($\EW>0.3$\AA)  relative to the host galaxy major axis is not consistent with being uniform at the $3.1$--$\sigma$ significance level. 
 The azimuth angle distribution is bi-modal
with about half the quasars aligned with the major axis  and the other half within $\alpha=30^{\circ}$ of the minor axis (Fig.~\ref{fig:bimodal}).
This bi-modal distribution confirms the presence of the azimuthal dependence of low-ionization gas around inclined disks as reported by \citet{BordoloiR_11a} at $z\sim1$ and by \citet{ChenY_10a} at $z\sim0$ and is inconsistent with the halo model of \citet{ChenHW_08a,TinkerJ_08a} and \citet{ChenHW_10a}.

  We associate the sight-lines aligned with the minor axis to sight-lines intercepting bi-conical outflows ('wind' sub-sample) and those  aligned with the major axis with sight-lines intercepting the outskirts  of galaxies ('disk' sub-sample). The dichotomy in azimuth angle is also present in the instantaneous SFR (Fig.~\ref{fig:SFR}). 

Using the 'wind' sub-sample, the data show that the outflows traced by low-ionization lines such as
\MgII\ have several properties:
\begin{itemize}
\item The bi-modal distribution of the $\alpha$ angle (Fig.~\ref{fig:bimodal})
shows that the outflows are rather well-collimated, covering a total solid angle $\Omega_w\simeq2$ accounting for both sides of the cone.
\item The wind speeds $V_{\rm out}$ inferred from the  \MgII\ absorption kinematics
and  a bi-conical wind model (Fig.~\ref{fig:J0814}--\ref{fig:J2250}) 
are of the order of the rotation speed.
\item The wind speeds tend to be smaller than (or equal to) the escape velocity, indicating that the low-ionization gas is not escaping the halo.  
\item The radial dependence of the \MgII\ equivalent width 
follows approximately the expected $b^{-1}$ dependence 
for pure geometry dilution with a scatter of 0.24~dex (Fig.~\ref{fig:radial}).
\item The mass outflow rates are about $2\times$ the current SFR, ranging from 1 to 6 \mpy\
using the wind speed and the empirical relation between \EW\ and \NHI\   \citep[e.g.][]{MenardB_09a}. 
Compared to the orders of magnitude uncertainties in
the best estimates from galaxy spectroscopy \citep[e.g.][]{HeckmanT_00a,PettiniM_02a},
our mass outflow rates are  accurate  to within $\sim50$\%, where most of the uncertainty
lies in the \NHI\ factor.
\end{itemize}

In Appendix~\ref{appendix:disk}, we show that 
our bi-conical outflows are consistent with the inclination dependence reported by \citet{KacprzakG_11b} if the azimuth angle is taken into account.
In particular, the scatter in the \EW-$b$ relation   is reduced \citep[as in][]{KacprzakG_11b}
when a correction of the type $X\propto1/\cos i$ is applied to the `disk' sub-sample.
On the other hand, this correction increases the scatter for the `wind' sub-sample, as 
one might have expected since this inclination correction is not appropriate in this case.
Interestingly,   the \EW-$b$ relation appears to be much steeper
($\propto b^{-3}$)
for the 'disk' sub-sample than for the 'wind' sub-sample
($\propto b^{-1}$) \citep[see also][]{ChurchillC_12a}.

Our results open a new and promising way to study the physical properties of
galactic outflows at high-redshifts using quasar absorption lines.
In the near future with larger samples, we will be able to investigate 
further  the properties of galactic outflows. 
In particular, a larger sample will allow us to test whether
 the loading factor 
 $\eta\equiv M_{\rm out}$/SFR is a function of circular velocity $V_c$,
as being assumed 
in numerical simulations by \citet{OppenheimerB_10a} and others.

\section*{Acknowledgments}
 We are very grateful to Dr. M. Fuller for running the Summer Research program at UC Santa Barbara.  We thank M. T. Murphy and S. Genel for stimulating discussions and 
their constructive comments on the draft.
   This work was partly supported by a Marie Curie International Outgoing
Fellowship (PIOF-GA-2009-236012) within the 7th European Community Framework Program. This work was supported in part by the National Science Foundation through grants AST-080816, AST-1109288 and AST-0708210.
Funding for the SDSS and SDSS-II has been provided by the Alfred P. Sloan Foundation, the Participating Institutions, the National Science Foundation, the U.S. Department of Energy, the National Aeronautics and Space Administration, the Japanese Monbukagakusho, the Max Planck Society, and the Higher Education Funding Council for England. The SDSS Web Site is http://www.sdss.org/.
 The SDSS is managed by the Astrophysical Research Consortium for the Participating Institutions. The Participating Institutions are the American Museum of Natural History, Astrophysical Institute Potsdam, University of Basel, University of Cambridge, Case Western Reserve University, University of Chicago, Drexel University, Fermilab, the Institute for Advanced Study, the Japan Participation Group, Johns Hopkins University, the Joint Institute for Nuclear Astrophysics, the Kavli Institute for Particle Astrophysics and Cosmology, the Korean Scientist Group, the Chinese Academy of Sciences (LAMOST), Los Alamos National Laboratory, the Max-Planck-Institute for Astronomy (MPIA), the Max-Planck-Institute for Astrophysics (MPA), New Mexico State University, Ohio State University, University of Pittsburgh, University of Portsmouth, Princeton University, the United States Naval Observatory, and the University of Washington.


\begin{table*}
\centering
\caption{Summary for galaxy-QSO pairs. \label{table:summary}}
\begin{tabular}{llllccccccccc}
\hline
QSO	 	& $z_{\rm abs}$ & Galaxy	&  $z_{\rm em}$ &   $W_r$(\AA) 	  &  $M_r$ & $V_{\rm max} \sin i$  & SFR$_{\Ha}$  &  $b$ & Ref.   \\   
(1)   		&	(2)	&	(3)	& (4)	&  (5)&  (6)  &  (7)	& (8) & (9) & (10)   \\
\hline
SDSSJ005244.23$-$005721.7 & 0.13460 & J005244G1 & 0.13426 &  1.46/1.23 &   -21.40 & 144 & 0.05 &  32.4 & K11  \\
SDSSJ081420.19$+$383408.3 & 0.09833 & J081420G1 & 0.09801 & 0.57/0.28 &  -20.13 & 131 & 1.27 & 51.1 & K11  \\
SDSSJ091119.16$+$031152.9 & 0.09636 & J091119G1 & 0.09616 & 0.82/0.34 &  -20.98 & 231 & 0.26 & 71.2 & K11 \\
SDSSJ092300.67$+$075108.2 & 0.10423 & J092300G1 & 0.10385 & 2.25/1.40 &  -21.58 & 108 & 0.02 & 11.9 & K11 \\
SDSSJ102847.00$+$391800.4 & 0.11411 & J102847G1 & 0.11348 & 0.30/0.13 &  -20.22 &162 & 3.75 & 89.8 & K11\\ 
SDSSJ111850.13$-$002100.7 & 0.13158 & J111850G1 & 0.13159 & 1.93/1.82 &  -20.40 & 116 & 1.96 & 25.1 & K11  \\
SDSSJ114518.47$+$451601.4 & 0.13402 & J114518G1 & 0.13389 & 1.06/1.07 &  -21.21 & 162 & 2.59 & 39.4 & K11 \\
SDSSJ114803.17$+$565411.5 & 0.10433 & J114803G1 & 0.10451 & 1.59/1.25 &  -21.58 & 67 & 0.15 & 29.1 & K11\\
SDSSJ144033.82$+$044830.9 & 0.11307 & J144033G2 & 0.11271 & 1.18/0.93 &  -20.22 & 112 & $\cdots$ & 24.9 & K11 \\
SDSSJ161940.56$+$254323.0 & 0.12501 & J161940G1 & 0.12438 & 0.32/0.28 &  -21.09 & 74 & 0.06 & 45.7 & K11\\
SDSSJ225036.72$+$000759.4 & 0.14837 & J225036G1 & 0.14826 & 1.08/1.11 &  -21.47 & 240 & 1.36  & 53.9 & K11\\
\hline
\end{tabular}		\\
(1) Quasar name; (2) \MgII\ absorption redshift; (3) Galaxy name; (4) Galaxy spectroscopic redshift;  (5) Equivalent widths for \MgII2796 and 2803\AA; 
(6) Absolute magnitude; (7) Observed rotation curve velocity (\kms); (8) SFR in \mpy; (9)
Impact parameter in kpc; (10) Reference:  K11 is for \citet{KacprzakG_11a}. 
\end{table*}

\begin{table*}
\caption{PA and $|\alpha|$ measurements.\label{table:measurements}}
\begin{tabular}{lcccrcccccccc}
\hline
Galaxy& $i$    & $i$ & PA   &  $|\alpha|$   & Class & SFR$_{\Ha}$ & $g-r$ &    $V_{\rm out}$  & $\dot M_{\rm out}$& $t_w$ \\
		&   K11 & Fit &   Manual/Fit  & Manual/Fit  &  &    (\mpy) &  & (\kms) & (\mpy) &  (Myr)   \\
(1) & (2) & (3) & (4) & (5)&  (6)  &  (7)	& (8) & (9)  & (10)  & (11) \\
\hline
& & & & $\alpha>60$ \\
J081420G1  & 40$\pm$2 & 35$\pm$2  & 30/18$\pm$2 &   79/67$\pm$2   &   Wind  & 1.27   & 0.57 &   175$\pm$25  & 2.2$\pm$1.1 &  290  \\
J091119G1  & 82$\pm$2 & 75$\pm$2  & $\cdots$/53$\pm$1   &    65/63$\pm$1  &   Wind  & 0.26	 & 1.02	&   500$\pm$100 & 6.8$\pm$3.4 & 140  \\ 
J092300G1  & 56$\pm$2 & 41$\pm$1  & 20/20$\pm$2 &    82/84$\pm$2  & Ambig.     & 0.02 &	1.2	&   200(?)   & 1.4$\pm$0.7 &  n.a.\\
J102847G1  & 54$\pm$2 & 49$\pm$2  & 89/90$\pm$5 &    76/83$\pm$5  &   Wind  & 3.75   & 0.57  &    300$\pm$25  & 1.0$\pm$0.5 &  250  \\
J111850G1  & 30$\pm$2 & 34$\pm$1  & 86/85$\pm$5 &    86/59$\pm$5  &   Wind  & 1.96   & 0.8 	&  175$\pm$80  & 6.0$\pm$3.0 & 140  \\
J225036G1  & 70$\pm$2 & 69$\pm$3  & 56/65$\pm$1 &    77/69$\pm$1  &   Wind  & 1.36   & 1.1    &  225$\pm$50  & 2.2$\pm$1.1 & 250   \\
 & & & $\alpha<20$ \\
J005244G1  & 45$\pm$5 & 42$\pm$2  & 40/43$\pm$2 &    20/22$\pm$2  &   `Disk'  & 0.05  & 1.0	&   n.a.&   n.a. & n.a. \\
J114518G1 & 34$\pm$2 & 34$\pm$1  & 44/38$\pm$5   &    15/21$\pm$5   & Ambig.  & 2.59  &	0.67 &   125$\pm$25  &   n.a. & n.a. \\
J114803G1  & 45$\pm$3 & 39$\pm$1  & 27/31$\pm$2 &    8/10$\pm$2   &   `Disk'  & 0.15   & 1.04	&   n.a.&  n.a. & n.a.  \\
J144033G2  & 55$\pm$5 & 45$\pm$2  & 69/75$\pm$4 &    7/13$\pm$4   &   `Disk'  & $\cdots$ & 0.50&   n.a.& n.a. & n.a.\\
\hline
\hline
J161940G1  & 12$\pm$12 & 5$\pm$20 & 7/-64$\pm$65 & unconstr. & n.a. &  $\cdots$  & 1.0   &   n.a. & n.a. & n.a. & \\
\hline
\end{tabular}\\
(1) Galaxy name;
(2) Galaxy inclination $i$ (degrees) from \citet{KacprzakG_11a} who used a bulge$+$disk decomposition;
(3) Galaxy inclination $i$ (degrees) from a one component S\'ersic fit;
(4) Galaxy position angle (PA) (degrees) measured manually or from our 2D fits;
(5) Azimuth angle $|\alpha|$ of the quasar location with respect to the galaxy major axis;
(6) Classification of the quasar-galaxy pair. `Wind' refers to sight-lines
whose \MgII\ kinematics can be explained with our model. `Disk' refers to 
sight-lines whose \MgII\ kinematics are likely related to some other physical process taking place in connection with the major axis.
(7) {\it Instantaneous} SFR in \mpy\ derived from \Ha\ taken from \citet{KacprzakG_11a}
assuming a Salpeter IMF and no reddening;
(8) $g-r$ color;
(9) Radial outflow speed  in \kms\ inferred from the \MgII\ kinematics;
(10) Mass outflow rates in \mpy\ derived from Eq.~\ref{eq:outflow};
(11) Travel time in Myr from the galaxy to the observed impact parameter  ($b/V_{\rm out}$).
\end{table*}
 

\appendix

\section{The kinematics of the Disk Sub-sample}

\label{appendix:disk}

In the main body of this paper, 
we find that a significant fraction of the \MgII\ absorbers with \EW\ from 0.5 to 3\AA\ are
found along the minor axis, i.e. are not co-planar with the galaxy host.
On the other hand, \citet{KacprzakG_11b} argued that \MgII\ absorbers are co-planar  based
on their finding that   the disk inclination influences the scatter of the \EW--$b$ relation 
for a sample of $z\sim0.5$ QSO-galaxy pairs.
Thus, their results appear to be incompatible with our conical winds since they both apply to the same \EW\ range of 0.5--3\AA.  

Here, we attempt to reconcile these two results. In particular,  we return to the 'disk' sub-sample with ($|\alpha|<45^{\circ}$) since an inclination effect should be present predominantly for this sub-sample. For extended gaseous disks,
 we expect that \EW\  is related to the path length $X$ intercepted by the QSO sight-line.
Because there are only three pairs in the \citet{BartonE_09a} and \citet{KacprzakG_11a} sample that meet the $\alpha$ criteria, we include pairs from  the $z\sim0.5$  \citet{KacprzakG_11b}  sample
using the same criteria ($|\alpha|<45^{\circ}$), excluding pairs where the   uncertainty in the azimuth angle  is greater than 30$^{\circ}$ (3$\sigma$). 

\begin{figure*}
\includegraphics[width=8.5cm]{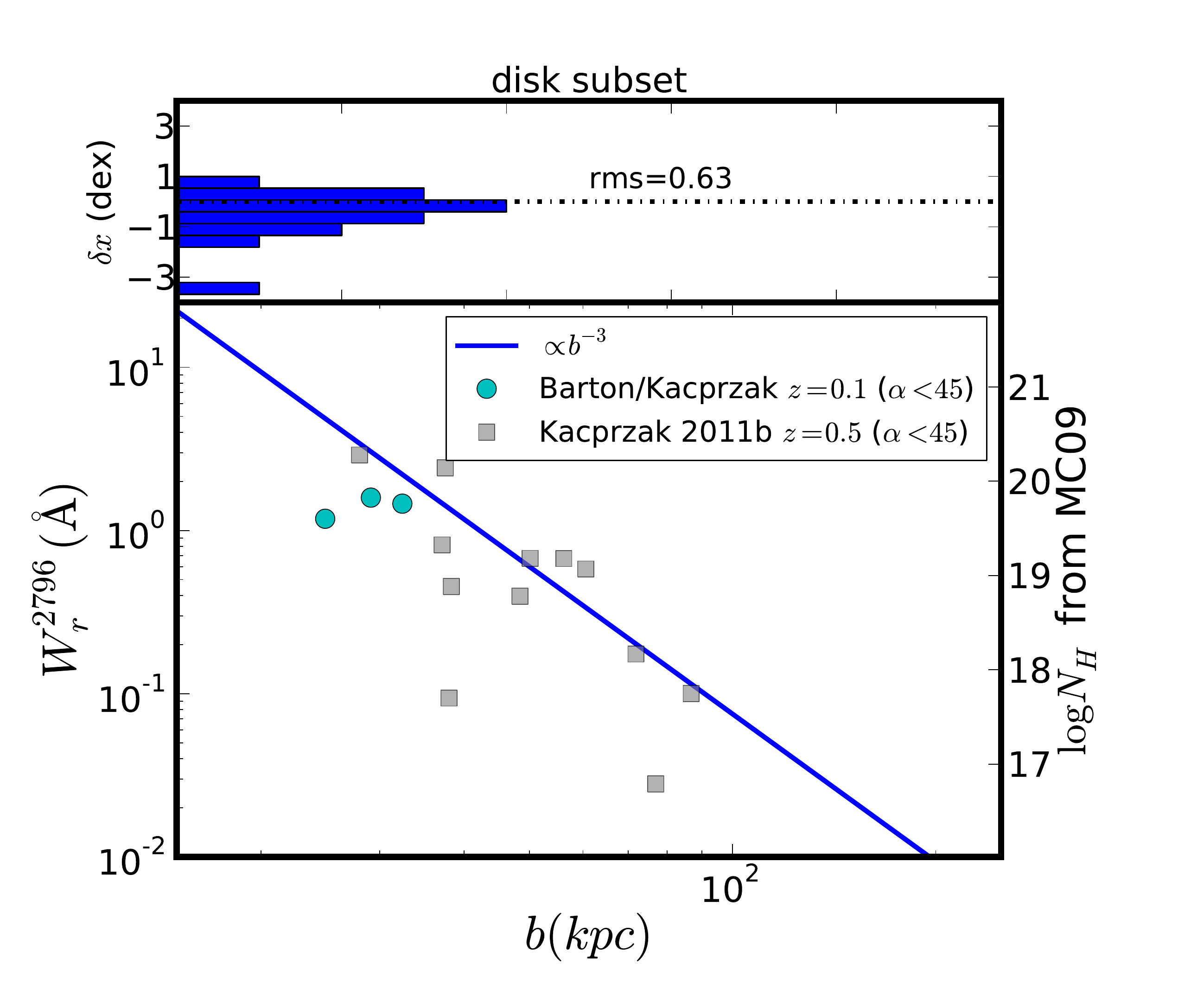}
\includegraphics[width=8.5cm]{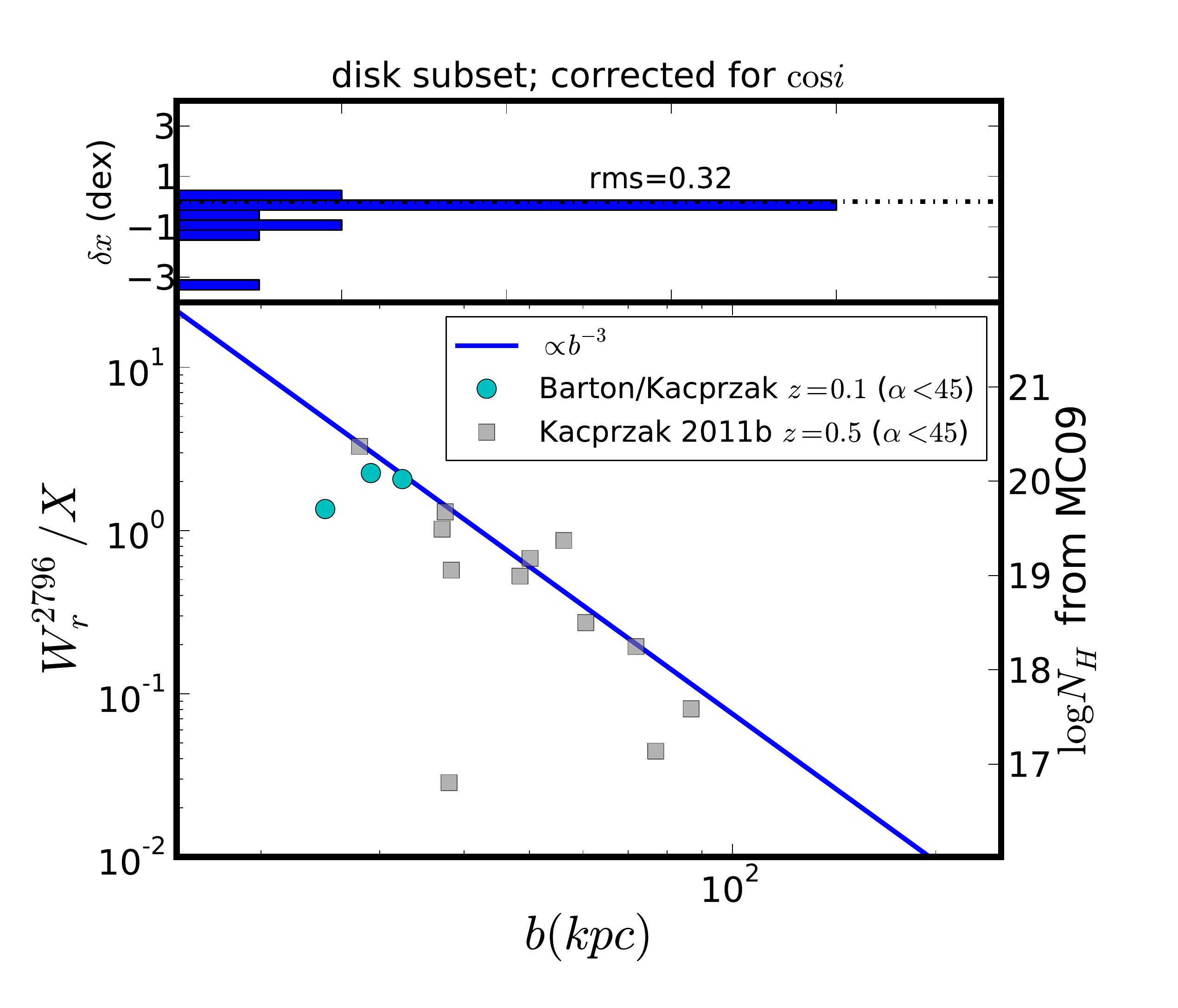}
\caption{{\bf (a):} \EW\ as a function of impact parameter $b$ for QSO-galaxy pairs 
classified as `disk', i.e. with $|\alpha|<45$.
{\bf (b):} Same as (a) with the $\EW$ normalized to the disk path length $X=X_0/\cos i$, where $i$ is the galaxy inclination. 
The top panels show that the scatter is reduced from 0.63 dex to 0.32 dex. 
This shows that such QSO-galaxy pairs near inclined disks with $|\alpha|\sim0$ are intercepting
either the extended parts of gaseous disks or, as argued in \citet{KacprzakG_11b}, 
  the accretion material that can also be co-planar according to the recent simulations of \citet{StewartK_11a}.}
\label{fig:disk}
\end{figure*}

Figure~\ref{fig:disk}(a) shows the \EW\ as a function of impact parameter $b$ for the $z=0.1$ ($z=0.5$) QSO-galaxy pairs shown as the cyan circles (squares) respectively. 
The solid line shows a fiducial $\propto b^{-3}$ radial dependence. 
The top panel shows that the residual scatter from this relation are $\sim0.63$~dex.

For this 'disk' sub-sample, we  expect that the absorption equivalent width \EW\  is related to the path length $X$ intercepted by the QSO sight-line.
For a simple slab geometry, the path length $X$ is expected to be  $\propto1/\cos i$ where $i$ is the disk inclination.  Hence, one would expect that the observed \EW\ is 
\begin{equation}
\EW = \tilde W_r^{\lambda 2796}\cdot X,
\end{equation} 
where $\tilde W_r$ is the equivalent width for face-on disks. 
 Figure~\ref{fig:disk}(b) shows the \EW-$b$ relation for the  
equivalent width $\tilde W_r$ corrected to a face-on inclination using $X=X_0/\cos i$.
We set $X_0$ to 0.5 corresponding to an averaged inclination of 60$^{\circ}$ given that the average inclination is $<i>=57^{\circ}$ \citep[see the appendix of][]{LawD_09a}. 
The residual scatter in the \EW--$b$ relation is significantly reduced to 0.32~dex, as indicated by the top histogram. This factor of 2 improvement in the scatter rms shows that $\cos i$ plays a large role
for the \EW\ of inclined disks, as stated in \citet{KacprzakG_11b}.

 This exercise shows that, in some cases, the absorption is co-planar and coupled to the galaxy inclination \citep{KacprzakG_11b}.  As stated in 
\citet{KacprzakG_11b}, 
the absorbing material could also be tracing the accretion of baryons
since according to \citet{StewartK_11a} such infalling material is predominantly coupled to the galaxy
angular momentum and might dominate the \MgII\ cross-section.

\begin{figure*}
\includegraphics[width=8.5cm]{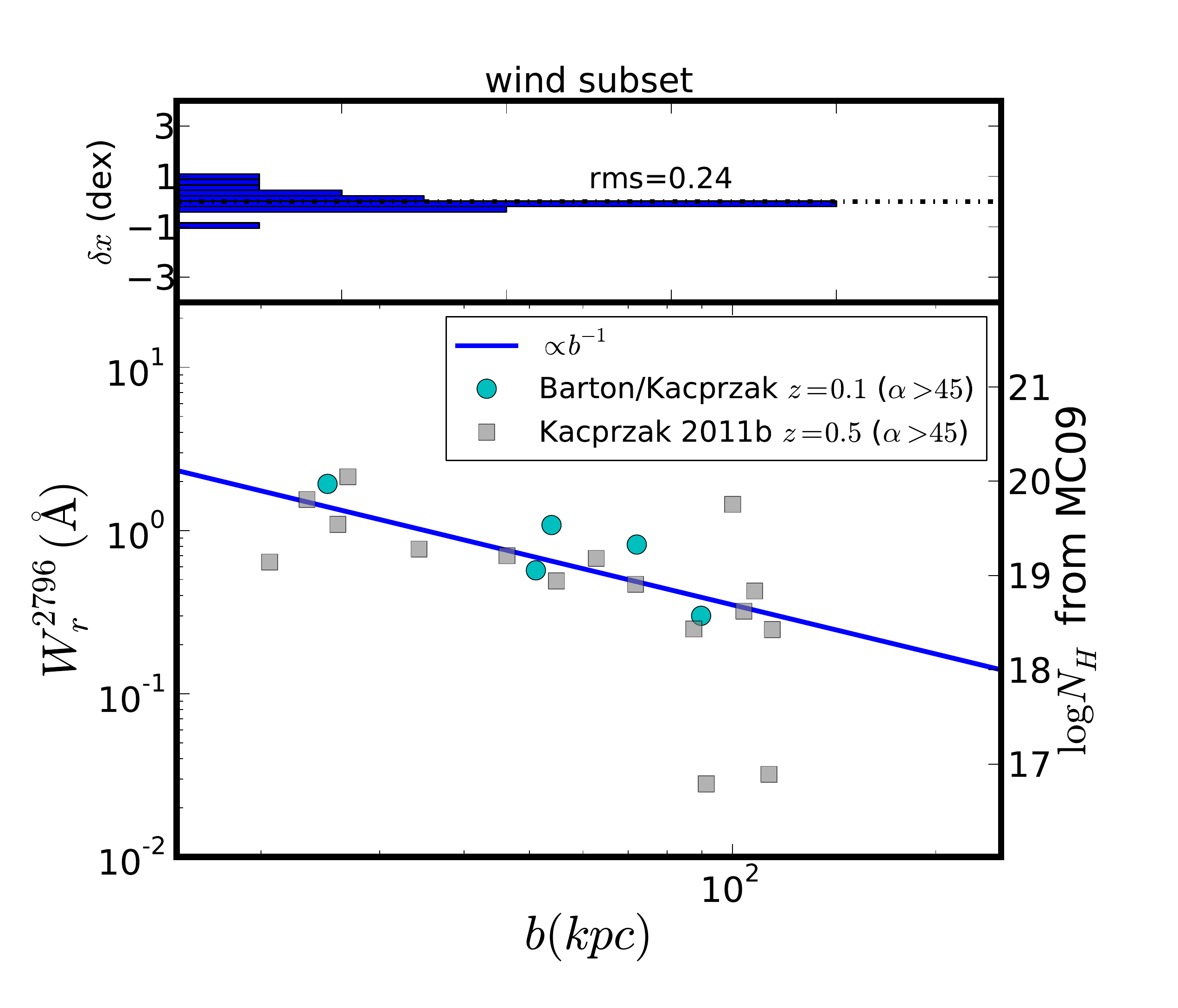}
\includegraphics[width=8.5cm]{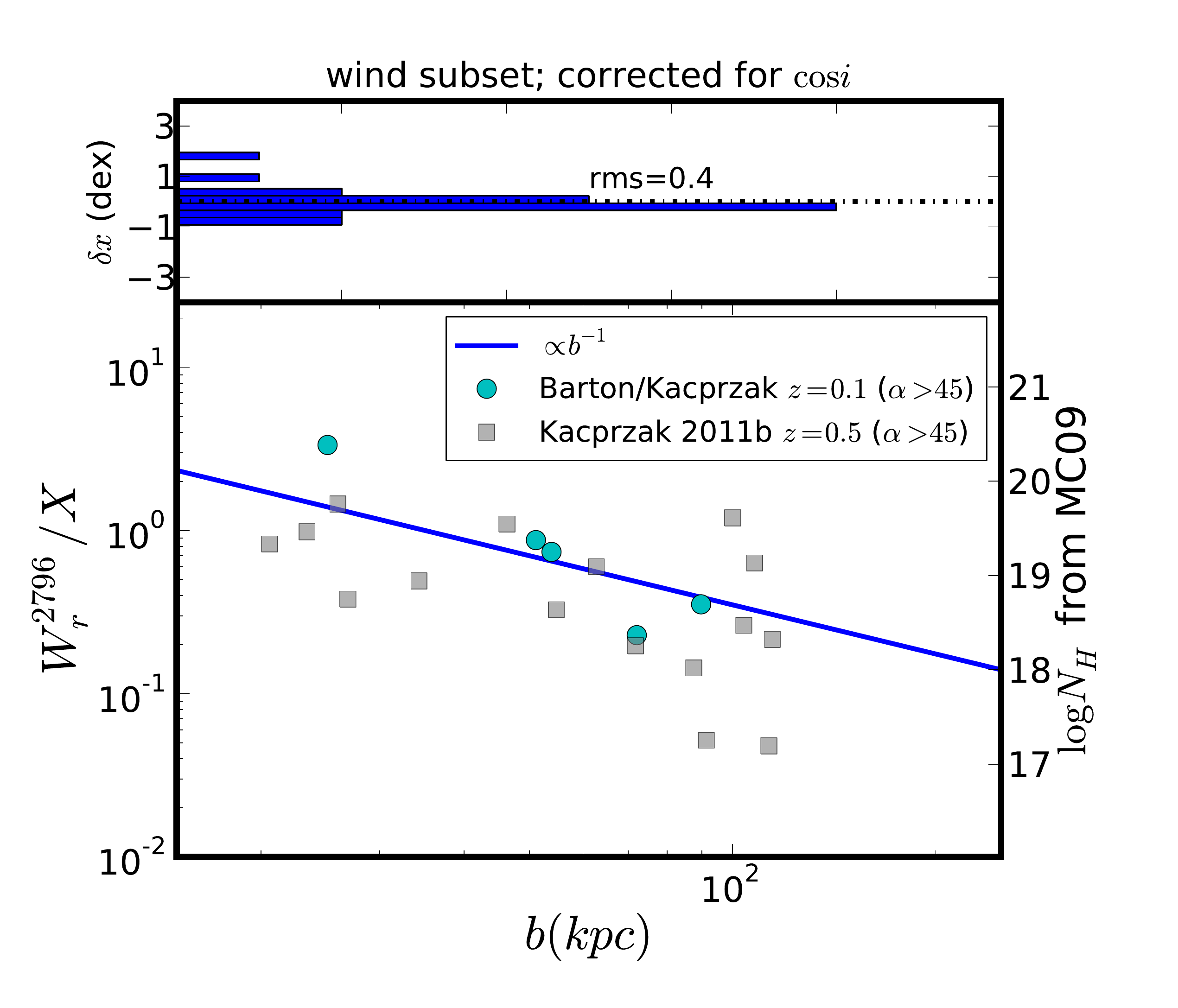}
\caption{{\bf (a):} \EW\ as a function of impact parameter $b$ for QSO-galaxy pairs 
classified as `wind', i.e. with $|\alpha|>45$.
{\bf (b):} Same as (a) with the $\EW$ normalized to the disk path length $X=X_0/\cos i$, where $i$ is the galaxy inclination. 
The top panels show that the scatter is {\it increased} from 0.24 dex to 0.40 dex. 
This shows that the disk path length $X$  is not appropriate for
QSO-galaxy pairs classified as 'wind', as one might have expected.}
\label{fig:windy}
\end{figure*}

Conversely, for the 'wind' sub-sample presented in section~\ref{section:radial},
 the same path-length correction should not apply and as a consequence it should {\it increase} the scatter in the \EW--$b$ relation. 
Figure~\ref{fig:windy} shows the \EW--$b$ relation for the `wind' sub-sample
uncorrected (left) and corrected for the disk path length (right).
This figure shows that the scatter increases from 0.24~dex to 0.4~dex and  demonstrates that the $X$ path-length is not applicable to this subset of \MgII\ absorbers.
Hence, our bi-polar outflows and the planar effects reported by  \citet{KacprzakG_11b} are not inconsistent with each other and this exercise demonstrates the importance of the azimuth angle in 
interpreting \MgII-galaxy pairs.

\section{Mass outflow rates}
\label{appendix:outflow}

Given that we are using quasar absorption lines to determine mass outflow rates $\dot M_{\rm out}$
  for the first time, we show all the steps in deriving the $\dot M_{\rm out}$ equation  used in this paper (Eq.~\ref{eq:outflow}).

In most general terms, the outflow rate $\dot M_{\rm out}$  for a fluid moving at a velocity $\mathbf V$ through an area $\Omega$ is
\begin{eqnarray}
\dot M_{\rm out}(r)&\equiv& \int_{\Omega} {\rm d}A\; \rho(r) \;\mathbf V\cdot\hat\mathbf n,\label{eq:generic}
\end{eqnarray}
where $\hat\mathbf n$ is the normal to the surface.
For a cone of opening angle $\theta_{\rm max}$,  the outflow speed is normal to $A$ in spherical coordinate, and this reduces to $\dot M_{\rm out}= \rho_0\;r_0^2\;V_{\rm out}\;\Omega$, where $\Omega=2\pi(1-\cos \theta_{\rm max})$.
Because,   the gas column density of a radial sight-line is $N\equiv\int_{r_0}{\rm d}r \;\rho(r)=\rho_0\,r_0$
  for a fluid  obeying the continuity equation ($\rho(r)r^2$=const), the outflow rate reduces to the
trivial equation $\dot M_{\rm out}=N\,r_0\,V_{\rm out}\Omega$ \citep{HeckmanT_00a,MartinC_05a}.
In the case of a conical geometry, with a transverse sight-line at impact parameter $b$, the outflow rate reduces
to a similar form  
\begin{equation}
\dot M_{\rm out}\propto N(b)\,b\,V_{\rm out}.
\end{equation}

For a transverse sight-line intercepting the symmetric
$ z$-axis  at $b=b_z$ of a cone, the integral in Eq.~\ref{eq:generic} is performed
on  the cross-section $A$ of the cone at $b_z$.  Using $t$ as the radius on
the cross-section $A$, the velocity $V_z$
normal to $A$ is $V_z=\mathbf V\cdot \hat\mathbf n=V_{\rm out}\frac{b_z}{\sqrt{b_z^2+t^2}}$.
Hence, the outflow rate $\dot M_{\rm out}(b_z)$ is $\int_{A}  {\rm d}A \; \rho(r) \;V_z$, i.e.
\begin{eqnarray}
\dot M_{\rm out}(b_z)&=& \rho_0\,r_0^2\int_{0}^{r_m}2\pi\,t{\rm d}t\; \frac{1}{b_z^2+t^2}V_{\rm out}\frac{b_z}{\sqrt{b_z^2+t^2}}\nn\\
&=&\rho_0\,r_0^2\;2\pi\;b_z\;V_{\rm out}\int_{0}^{r_m}{\rm d}t\; \frac{t}{(b_z^2+t^2)^{3/2}}
\end{eqnarray}•
where  $t$ is  bound to a maximum $r_m=b_z\tan\theta_{\rm max}$ by the cone edge.
After integration, we find
\begin{eqnarray}
\dot M_{\rm out}(b_z)&=& \rho_0\,r_0^2\;V_{\rm out}\;2\pi[1-\cos\theta_{\rm max}] \label{eq:Moutflow}\\
&\simeq & \rho_0\,r_0^2\;V_{\rm out}\pi\theta_{\rm max}^2.\nn
\end{eqnarray}

The column density $N(b)$ for a transverse sight-line intercepting the symmetric
$ z$-axis at $b=b_z$ is 
\begin{eqnarray}
N(b)&=&\rho_0\,r_0^2\int_{-x_1}^{x_1}{\rm d}x\frac{1}{b^2+x^2} \nn\\
&=&\frac{\rho_0\,r_0^2}{b}\left.\arctan \frac{x}{b}\right|_{-x_1}^{x_1}=\frac{\rho_0\,r_0^2}{b}2\theta_{\rm max}
\label{eq:Ntrans}
\end{eqnarray}
since the opening angle $\theta_{\rm max}$ defines the integration range
$x_1=b \tan\theta_{\rm max}$.

In the most general case, for a transverse sight-line that is offseted  from the $ z$-axis
by $b_y$,   where  $x_1=\sqrt{\tan^2\theta_{\rm max}b_z^2-b_y^2}$, and $b=\sqrt{b_y^2+b_z^2}$,
we have the column density $N(b)$
\begin{eqnarray}
N(b_z)&=&\frac{\rho_0\,r_0^2}{b}2\arctan\frac{\sqrt{\tan^2\theta_{\rm max}b_z^2-b_y^2}}{\sqrt{b_z^2+b_y^2}}
\end{eqnarray}
which reduces to Eq.~\ref{eq:Ntrans} when $b_y=0$.

Combining Eq.~\ref{eq:Moutflow} with Eq.~\ref{eq:Ntrans}, we have that the outflow rate determined
from transverse sight-lines is:
\begin{eqnarray}
\dot M_{\rm out}(b)&=&\frac{N(b)\,b}{2\theta_{\rm max}}V_{\rm out} 2\pi[1-\cos\theta_{\rm max}] \\
&\simeq& N(b)\,b\,V_{\rm out}\frac{\pi}{2}\theta_{\rm max}\label{eq:Moutflow:transverse}
\end{eqnarray}•

\bsp

\label{lastpage}

\end{document}